\documentclass[preprint,12pt]{elsarticle}
\biboptions{sort} 

\usepackage[utf8]{inputenc}
\usepackage{amssymb}
\usepackage[figuresright]{rotating}
\usepackage{booktabs}
\usepackage{multirow}
\usepackage{tabularx}
\usepackage{caption}
\usepackage{subcaption}
\usepackage{hyperref}
\usepackage{tabularray}
\usepackage{longtable}
\usepackage{makecell}  
\usepackage{longtable} 
\usepackage{tcolorbox}  

\usepackage{amssymb}
\usepackage{amsmath}

\newcommand*{\rqOne}{How much software engineering research on Jupyter Notebook has been published?}
\newcommand*{\rqTwo}{Which software engineering topics are being studied in software engineering research on Jupyter Notebook?}

\journal{Journal of Systems and Software}

\begin{document}

\begin{frontmatter}



\title{A Systematic Literature Review of Software Engineering Research on Jupyter Notebook}


\author[ualberta]{Md Saeed Siddik} 


\author[queens]{Hao Li} 


\author[ualberta]{Cor-Paul Bezemer} 

\affiliation[ualberta]{organization={University of Alberta},
            city={Edmonton},
            country={Canada}}
\affiliation[queens]{organization={Queen’s University},
            city={Kingston},
            country={Canada }}
            
\begin{abstract}
\textit{Context}:
Jupyter Notebook has emerged as a versatile tool that transforms how researchers, developers, and data scientists conduct and communicate their work. As the adoption of Jupyter notebooks continues to rise, so does the interest from the software engineering research community in improving the software engineering practices for Jupyter notebooks. 

\textit{Objective}: 
The purpose of this study is to analyze trends, gaps, and methodologies used in software engineering research on Jupyter notebooks. 

\textit{Method}: 
We selected 199 relevant publications up to September 2025, following established systematic literature review guidelines. We explored publication trends, categorized them based on software engineering topics, and reported findings based on those topics.

\textit{Results}: 
The most popular venues for publishing software engineering research on Jupyter notebooks are related to human-computer interaction instead of traditional software engineering venues. 
Researchers have addressed a wide range of software engineering topics on notebooks, such as code reuse, readability, and execution environment. Although reusability is one of the research topics for Jupyter notebooks, only 82 of the 199 studies can be reused based on their provided URLs. Additionally, most replication packages are not hosted on permanent repositories for long-term availability and adherence to open science principles. 

\textit{Conclusion}: 
Solutions specific to notebooks for software engineering issues, including testing, refactoring, and documentation, are underexplored. Future research opportunities exist in automatic testing frameworks, refactoring clones between notebooks, and generating group documentation for coherent code cells. 
\end{abstract}


\begin{highlights}
\item This research provides the first comprehensive systematic literature review on software engineering research specifically targeting Jupyter notebooks, identifying 199 primary studies published up to September 2025 and categorizing them into 11 core software engineering topics.

\item This research reveals that a large portion of the studies have been published outside traditional software engineering venues, with Human-Computer Interaction conferences like ACM Conference on Human Factors in Computing Systems~(CHI) being the top publishing venues, highlighting the interdisciplinary nature of Jupyter Notebook research.

\item This research identifies a reusability gap in existing research, showing that only 82 out of 199 studies offer usable replication packages, and most are hosted on GitHub instead of permanent repositories, which violates open science best practices.

\item This research identifies that notebook-specific solutions for software engineering issues such as testing, refactoring, and documentation are relatively underexplored. Future directions include resolving duplicated execution numbers, refactoring inter-notebook clones, and generating grouped documentation for coherent-code cells are future directions derived from our study. 
    
\item This research proposes the integration of modern AI-based solutions into Jupyter notebooks to support various software engineering topics, including code search and code generation. Additionally, future research should leverage advanced AI techniques~(e.g., large language models), to improve conversational AI-powered assistants for automated code generation by multi-step workflow automation in data science notebooks.

\item Although the paper exceeds the recommended length due to the inclusion of detailed tables, figures, and categorized analyses (covering 11 topics and 21 subtopics), we believe that this extended content is essential for clearly and completely reporting our findings. As the first systematic literature review in this domain, we have carefully structured the paper to ensure readability. We believe the length is justified by the value and breadth of this paper's contributions.

\end{highlights}


\end{frontmatter}





\section{Introduction}
\label{sec:slr:intro}
In recent years, Jupyter Notebook has emerged as a powerful and versatile computing platform for data science, scientific research, and software development~\citep{perkel2018jupyter, rule2018exploration}. Its user-friendly interface has revolutionized the way researchers, developers, and data scientists conduct and communicate their work. Jupyter Notebook provides a unique platform for integrating code, visualizations, and narrative text, fostering reproducibility, and facilitating collaborative exploration of data and algorithms~\citep{wang2019data}. Seamlessly incorporating code, data, and output into a single file makes the Jupyter Notebook ideal for data analysis, scientific computing, and machine learning~(ML) tasks. Moreover, its support for multiple programming languages, including Python, R, and Julia, makes it versatile and widely accessible to a diverse community of researchers, data scientists, and educators. However, despite their popularity, notebooks have been associated with several challenges, such as low reproducibility rates, problems with their execution environments, and excessive code duplication~\citep{grotov2022large}.

As Jupyter notebooks are often created and maintained by users without a software engineering background~\citep{simmons2020large}, there has been a growing interest from software engineering researchers to help notebook users integrate best software engineering practices into their notebooks. As a result, there has been a growing body of software engineering research that targets Jupyter Notebook. In this paper, we conduct a systematic literature review (SLR) of such research and the problems addressed by it. 

We systematically searched academic publications to identify relevant studies related to software engineering research on Jupyter Notebook. We followed Kitchenham's guidelines~\citep{kitchenham2004procedures} for our SLR to ensure rigor and impartiality. We comprehensively searched all the papers on Jupyter Notebook indexed in the ACM digital library, IEEE Xplore, Elsevier, Springer Link, and DBLP databases published until September 2025. Then, we filtered the papers focused on software engineering research on Jupyter notebooks. The review process involved several steps, including screening the titles and abstracts of the identified studies, assessing their relevance based on the inclusion criteria, and extracting relevant data from the selected studies. We finally selected 199 papers as our primary studies. The overall characterization of our review based on Cooper’s taxonomy~\citep{Cooper1988Taxonomy} is shown in Table~\ref{tab:cooper}.
During our SLR, we focused on the following two research questions (RQs):

\begin{table}[t]
\scriptsize
\centering
\caption{Characterization of this SLR based on Cooper’s taxonomy~\citep{Cooper1988Taxonomy}.}
\label{tab:cooper}
\begin{tabular}{lp{11cm}}
\toprule
\textbf{Dimension} & \textbf{Our SLR} \\
\midrule
Focus & Research outcomes, research methods, practices or applications \\
Goal & Integration, identification of central issues \\
Perspective & Neutral representation \\
Coverage & Exhaustive review with selective citation (``Jupyter'' or ``notebook'' in the title) and limited to software engineering research \\
Organization & Conceptual (topic-centric; see Table~\ref{tab:SEtopics}) \\
Audience & Specialized scholars in software engineering, practitioners \\
\bottomrule
\end{tabular}
\end{table}

\begin{itemize}
    \item \textbf{RQ1: \rqOne} We analyzed current trends in the publication of software engineering research on notebooks to help future researchers identify potential venues for their work. The number of publications has gradually increased over the years. 
    Most (104 of 199 studies) software engineering research on notebooks has been published at conferences, indicating a fast-moving field. Our research found that other venues beyond core software engineering conferences published the highest number of studies; for example, the ACM Conference on Human Factors in Computing Systems (CHI) published 20 studies. Conversely, the International Conference on Software Engineering (ICSE) published seven studies on software engineering research on notebooks. According to our study, a significant number of the articles we examined, specifically 37 out of 199 (18.8\%), had affiliations with industry institutions such as Microsoft Research (13 studies) or IBM Research Lab (6 studies). This highlights the practical importance and real-world applications of the research findings.
    We identified 87 studies that provided URLs to replication packages. Of these, 82 URLs were deemed reusable, while 5 were not, due to missing source code or unclear execution instructions. Most replication packages (69 studies) are hosted in GitHub repositories, which is against the best practices for open science in software engineering~\citep{Mendez_2020} as such repositories can disappear over time. Only 17 replication packages were hosted in a permanent repository, such as Zenodo or Figshare.

    \item \textbf{RQ2: \rqTwo} 
    We categorized the primary studies into 11 software engineering topics: code reuse and provenance, managing the computational environment and workflow, readability of notebooks, datasets of notebooks, documentation of notebooks, testing and debugging, visualization in notebooks, best practices, cell execution order, notebook code generation, and supporting other programming paradigms. Our findings indicate that managing computational environment and workflow is the most extensively researched topic related to Jupyter Notebook, with 42 studies focusing on it. This is followed by code reuse and provenance with 41 studies, and documentation of notebooks with 22 studies. These results suggest that researchers are primarily concerned with code cells in notebooks, which often require significant human effort to understand. We also identified 22 publicly available Jupyter Notebook datasets used in the studies, most of which were sourced from Kaggle and GitHub repositories.
\end{itemize}

Our systematic literature review reveals that software engineering research on Jupyter notebooks is an active research direction with diverse publication venues beyond core software engineering. Our research indicates that studies on notebooks are highly focused on dealing with code cells in notebooks, which require considerable human effort to understand and manage. Although there are techniques and tools for software engineering in Jupyter notebooks, such as automated refactoring tools, testing frameworks, and strong documentation practices, they are often insufficient. These present a chance for improvement, since better tools can enhance the usability and quality of projects in Jupyter notebooks. Future research can focus on improving these tools to support collaboration and reproducibility in scientific computing and data analysis.

The remainder of this paper is organized as follows. Section~\ref{sec:slr:jupyter} gives an overview of the Jupyter Notebook platform. Section~\ref{sec:slr:method} describes our methodology. Sections~\ref{sec:slr:res:rq1} and \ref{sec:slr:res:rq2} present the results of the research questions. Section~\ref{sec:slr:frd} presents the future research directions derived from this SLR. Section~\ref{sec:slr:threat} identifies the threats to validity and Section~\ref{sec:slr:conc} concludes the paper.

\section{The Jupyter Notebook Platform}
\label{sec:slr:jupyter}

Jupyter Notebook allows users to edit and execute code inside a web-based interface. Unlike traditional IDEs like PyCharm, Jupyter Notebook provides a cell-based interface that seamlessly integrates code with output and allows individual code cell execution in any order. In this section, we explore the platform's features and its utilization in various fields.

\subsection{Unique Features of Notebooks}
Jupyter Notebook enables users to manage code, documentation, and output into a single document. Users can execute any code cell from anywhere in the notebook; they are not bound to follow the code cell's order sequence. Two main features distinguish notebooks from a traditional source code file: the cell-based structure and execution order.

\begin{figure}[!ht]
\centering
\includegraphics[width=0.75\textwidth]{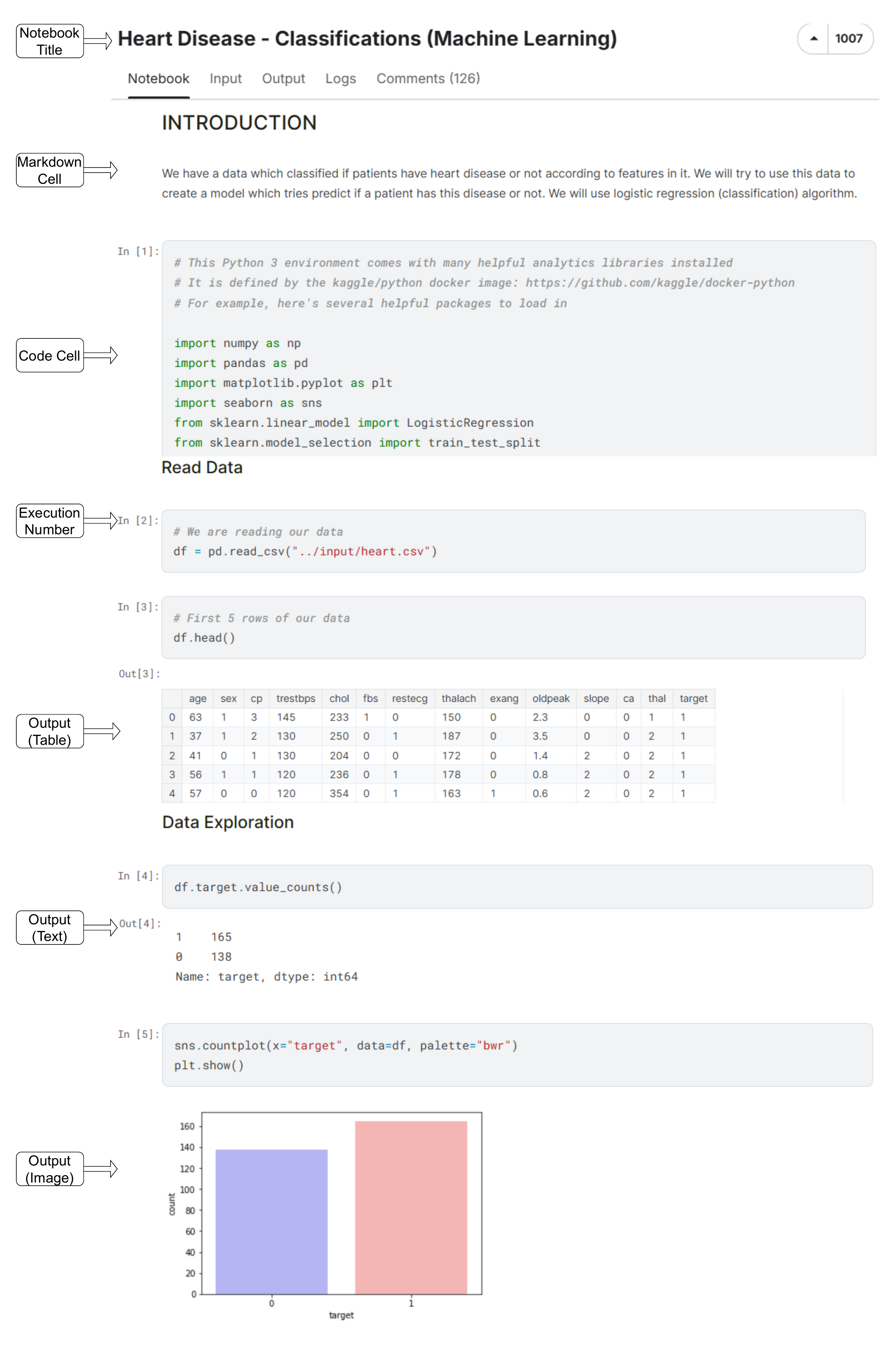}
\caption{An example of Jupyter Notebook's structure}
\label{fig:nbstructure}
\end{figure}

\subsubsection{Cell-Based Structure}
A Jupyter Notebook is composed of several types of cells: code cells for executing source code, Markdown cells for documentation, and raw cells that contain non-rendered contents. Furthermore, every code cell has an execution number, which records the order of cell execution, and an output cell that displays the result of the corresponding code cell. 
An illustrative example of this structure is shown in Figure~\ref{fig:nbstructure}, sourced from a publicly accessible notebook on Kaggle.\footnote{\url{https://www.kaggle.com/code/cdabakoglu/heart-disease-classifications-machine-learning}} Below is a detailed exploration of each cell type: 

\textbf{Code Cell:}
Code cells are the fundamental components of a Jupyter Notebook, where the source code (Python or other supported languages depending on the kernel) is written and executed. After execution, each code cell shows the results immediately below the cell. A notebook enables users to edit and execute each code cell independently, allowing them the flexibility to run cells in any order they prefer. This means that programmers are not obligated to execute their code sequentially from top to bottom or limited to making changes only at the end of the notebook. Instead, they can selectively modify and execute specific code cells at any point, fostering a more dynamic and interactive coding experience~\citep{chattopadhyay2020whats, grotov2022large}.

\textbf{Markdown Cell:}
Jupyter Notebook's Markdown cells provide the narrative to a notebook. They use plain text with Markdown syntax to create formatted text and media, including headings, lists, links, images, and formatted text to explain and guide the reader through the notebook's purpose, underlying logic, and methodology~\citep{pimentel2019large, venkatesh2021automated}. This aspect of notebooks is crucial for making them comprehensive and user-friendly, especially when sharing with peers who may not be familiar with the code. Comments in traditional programming are quite similar to this Markdown cell. However, a Markdown cell generally contains more information than traditional textual comments~\citep{liu2021haconvgnn, mondal2023cell2doc}.     

\textbf{Raw NBConvert Cell:}
A raw cell\footnote{\url{https://ipython.org/ipython-doc/3/notebook/nbformat.html\#raw-nbconvert-cells}} in a Jupyter notebook is a cell whose contents are included without modification when converted using nbconvert.\footnote{\url{https://ipython.org/ipython-doc/3/notebook/nbconvert.html\#nbconvert}} Unlike other cell types, raw cells are neither processed nor rendered within the notebook itself and are primarily intended for direct inclusion in exported outputs such as LaTeX documents or HTML files. This feature offers users flexibility in formatting and exporting notebook content for various publication or documentation purposes.

\textbf{Output Cell:}
The output of a code cell displays the results generated by the execution of the corresponding code cell. These can include textual data, tables, error messages, or visualizations created using libraries such as Matplotlib,\footnote{\url{https://pypi.org/project/matplotlib/}} Seaborn,\footnote{\url{https://pypi.org/project/seaborn/}} or Plotly.\footnote{\url{https://pypi.org/project/plotly/}} The outputs are automatically updated whenever the associated code cell is executed. This capability makes Jupyter notebooks especially valuable for data science and research, as outputs can immediately visualize and communicate results alongside code, enhancing understanding and analysis~\citep{raghunandan2023measuring, psallidas2022data}. 

\subsubsection{Execution Order}
The execution order in a Jupyter notebook refers to the sequence in which the code cells are executed. Unlike traditional scripts, code cells in a notebook can be executed non-linearly, depending on the user's needs and exploration path. The execution sequence is indicated by a number in brackets next to each code cell (e.g., \texttt{In [1]:}), which shows the most recent execution order (see Figure~\ref{fig:nbstructure} for the execution number). Jupyter Notebook allows users to execute any code cell multiple times or out-of-order. This flexibility allows notebook users to test specific parts of their code, troubleshoot issues, or make adjustments without rerunning the entire notebook~\citep{sato2022comparing, pimentel2021understanding, head2019managing}. This type of execution in notebooks also helps the iterative nature of exploratory data analysis~\citep{fangohr2019data}. However, out-of-order execution can also cause problems, such as failure to load dependencies if they are not properly managed in the code~\citep{wang2020restoring}. Researchers found that about 36\% of notebooks on GitHub did not execute in linear order~\citep{michael2022keeping}.

\subsection{Applications of Jupyter Notebooks}
Jupyter Notebook is a popular tool for data analysis and data science, often used for interactive exploration during experiments~\citep{perkel2018jupyter, kery2018story, werner2021bridging}. It is considered an effective tool for both experienced and novice data scientists~\citep{fruchart2022jupyter, chandel2022training}. Its flexibility in code execution facilitates short feedback cycles, which are essential for the iterative data analysis process~\citep{fangohr2019data}. While notebooks have been used with static visualizations, interactive visualizations can also be embedded and supported, as well as advanced visual analysis~\citep{ono2021interactive, wang2023supernova}. These allow users to quickly try out different data analysis options and observe the results with minimal effort. Notebooks are also used for generating personalized data narrations over a given dataset for interactive data exploration~\citep{chanson2022generating} and sharing data science work through presentation slides~\citep{zheng2022telling, wang2023slide4n, li2023notable, ouyang2024noteplayer}.

Jupyter notebooks are widely used in the education sector and in academic teaching~\citep{amoudi2023interactive, tan2021nascent, petersohn2023kopplung, casseau2023moon}. They provide an intuitive and user-friendly platform for teaching programming concepts and designing course structures~\citep{reades2020teaching, kastner2020teaching}. Educators use Jupyter notebooks to create coding tutorials, assignments, and lecture materials that foster active learning and engagement among students~\citep{amoudi2023interactive, al2022jupyter}. Research has shown that the use of Jupyter notebooks in educational settings significantly enhances the student's understanding of course materials~\citep{barba2019teaching}. Additionally, studies demonstrate that Jupyter notebooks are particularly convenient for teaching data science, as they eliminate the need for students to install software locally or use specific machines~\citep{van2020jupyter}. Despite their popularity for managing coding-based assignments, automatic assignment grading in Jupyter notebooks presents challenges. The standard automated assessment grading tool for Jupyter notebooks, such as \textit{nbgrader}~\citep{hamrick2016creating}) has limitations; for example, it cannot automatically submit scores to a Learning Management System (LMS) like Canvas and does not provide automated feedback to students~\citep{manzoor2020auto}. To address the automatic score submission, Malone et al. developed an automated assignment grading system for Jupyter notebooks that focuses on gamified cybersecurity exercises~\citep{malone2023securely}. 

Jupyter Notebook is popular for modeling and analyzing scientific tasks and performing experimental simulations~\citep{beg2021using, savira2021writing, penuela2021open, vandewalle2019integrating}. For example, Tran et al. developed a library for Jupyter Notebook that can set up and control additive manufacturing machines such as 3D printers~\citep{tran2023imprimer}. Wilsdorf et al. presented a Jupyter Notebook extension that lends support to modelers by automatically specifying and running suitable simulation experiments~\citep{wilsdorf2023nbsimgen}. Jupyter Notebook is also popular for analyzing GIS data to deal with big geospatial data~\citep{yin2017cybergis, yin2019cybergis, haedrich2023integrating}. For example, Valentine et al. implemented a data discovery studio for geoscience data discovery and exploration using Jupyter Notebook~\citep{valentine2021earthcube}. Yin et al. presented a cyberGIS framework to achieve data-intensive and scalable geospatial analytics using the Jupyter Notebook~\citep{yin2017cybergis}. Jupyter notebooks are also used as a toolbox for interactive surface water mapping~\citep{owusu2022pygee}, geospatial environmental data processing~\citep{terlych2021jupyter}, and astrophysical data-proximate analysis~\citep{juneau2021jupyter}. 

Notebooks are often used in the healthcare sector to analyze complex medical data~\citep{gruning2017jupyter, almugbel2018reproducible, ayobi2023computational}.  For example, Hao et al. from IBM Research developed customized healthcare data analysis pipelines in Jupyter Notebook to support healthcare users on user-friendly and reusable data analytics~\citep{hao2017developing}. Almugbel et al. developed a tool that allows users to easily distribute their biomedical data analysis through notebooks uploaded to a GitHub repository or a private server~\citep{almugbel2018reproducible}. The study provided four different Jupyter notebooks to infer differential gene expression, analyze cross-platform datasets, and process RNA sequence data. Biomedical researchers prefer notebooks to document and share their research with their community~\citep{almugbel2018reproducible, ayobi2023computational}. For example, researchers used Jupyter Notebook as a co-design tool that combines static illustrations and interactive ML model explanations to predict health risk in diabetes care~\citep{ayobi2023computational}. Furthermore, Llaunet et al. developed and shared a solution with centralized Jupyter Notebook code to support various medical applications such as medical image processing~\citep{launet2023federating}.
Beyond the above sectors, people use Jupyter Notebook in different areas. For example, journalists and media organizations use notebooks to analyze journalism content and visualize trends and patterns of newspaper data~\citep{showkat2021stories, stark2016towards}.

\section{Research Methodology}
\label{sec:slr:method}
\begin{figure}[t]
\centering
\includegraphics[width=0.75\textwidth]{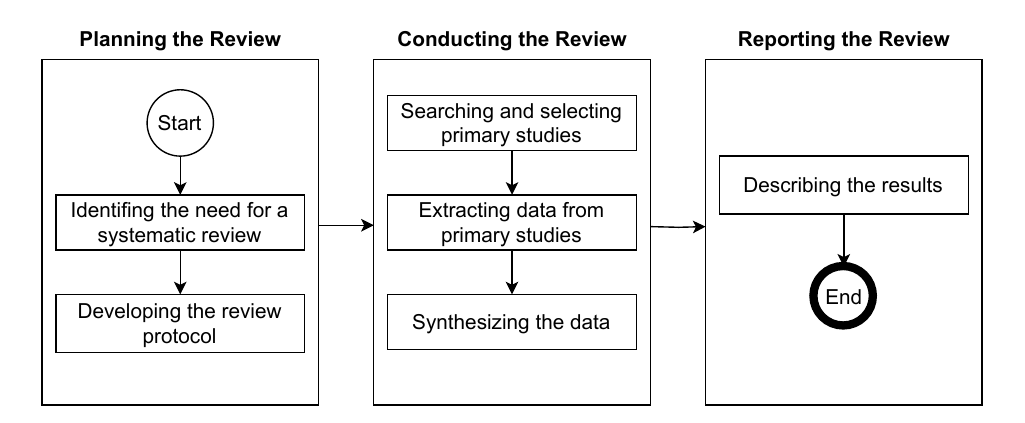}
\caption{The steps of our SLR on software engineering research on Jupyter Notebook based on Kitchenham’s guidelines~\citep{kitchenham2004procedures}}
\label{fig:slrsteps}
\end{figure}

For our systematic literature review of software engineering research on Jupyter notebooks, we followed the guidelines proposed by Kitchenham~\citep{kitchenham2004procedures}. The planning stage of our work includes two steps: (1)~identifying the need for a systematic review and (2)~developing the review protocol. In the conducting stage, based on the review protocol from the planning stage, we searched and selected the primary studies. We considered five different digital libraries (IEEE Xplore,\footnote{\url{https://ieeexplore.ieee.org/}} ACM Digital Library,\footnote{\url{https://dl.acm.org/}} Elsevier ScienceDirect,\footnote{\url{https://www.sciencedirect.com/}} Springer Nature Link\footnote{\url{https://link.springer.com/}} and DBLP\footnote{\url{https://dblp.org/}}) as the search space, as those cover all the major software engineering journals and conference proceedings published by renowned publishers. Then, we extracted and synthesized the data. We selected 199 papers as our primary studies after completing all the steps in this process.
Finally, in the reporting stage, we concluded the systematic review by reporting the collected data and findings. Figure~\ref{fig:slrsteps} summarizes the steps of our methodology. In this section, we explain each step in more detail.

\subsection{Planning the Review}

\subsubsection{Identifying the Need for a Systematic Review}
The need for a systematic review of software engineering research on Jupyter Notebook arises from the growing popularity and widespread adoption of Jupyter notebooks in various research fields. The widespread use of Jupyter notebooks has highlighted software engineering challenges such as poor code quality, lack of coding standards, and code duplication~\citep{adams2023comparison, wang2020better, koenzen2020code}. These difficulties have sparked notable interest among software engineering researchers in helping notebook users adopt software engineering practices in their work. As a result, more research has been conducted on these issues in Jupyter notebooks. This highlights the need to conduct a systematic literature review of this research and the challenges it addresses, as well as identify what should be studied next.

{
\scriptsize
\setlength\LTleft{0pt} 
\setlength\LTright{0pt}
\begin{longtable}{p{11.3cm}c}
\caption{Questions in the data extraction form} \label{tab:question} \\
\toprule
\textbf{Question (Q) - Description (D) - Rationale (R)} & \textbf{Target RQ} \\
\midrule
\endfirsthead

\toprule
\textbf{Question (Q) - Description (D) - Rationale (R)} & \textbf{Target RQ} \\
\midrule
\endhead

\bottomrule
\multicolumn{2}{r}{\textit{Continued on next page}} \\
\endfoot

\endlastfoot
Q: In which year was the study published?          & RQ1 \\ 
D: The publication year of the corresponding study. &   \\ 
R: The publication year helps to indicate the interest in this research topic across the years. &   \\
\midrule
Q: What is the publication type?         & RQ1 \\ 
D: The publication type of the study (\textit{Journal}, \textit{Conference}, \textit{Book}, \textit{Technical report}, or \textit{Other}). &   \\
R: Understanding the distribution of research across these publication types provides insights into the preferred channels for sharing knowledge in the field of software engineering research on Jupyter notebooks. &   \\
\midrule
Q: Was the research done with an industry collaborator?          & RQ1    \\ 
D: \textit{Yes} if any of the co-authors have an industry affiliation in the study, \textit{No} otherwise.  &   \\ 
R: Collaborations between academia and industry indicate that the research is conducted towards solving practical problems and has potential direct applications in industry settings. Understanding the extent of industry collaboration can help in assessing the balance between theoretical and applied research in software engineering research on Jupyter notebooks. &  \\
\midrule
Q: What type of solution has been proposed?        & RQ1    \\ 
D: The type of proposed solution (\textit{theoretical framework}, \textit{developed solution}, or \textit{empirical analysis}).    &   \\ 
R: Identifying the type of solution proposed in each study helps to understand the nature of the study: knowing the type of solution provides insights into the focus of the research and its methodology - whether it is more conceptual, application-based, or data-driven. & \\
\midrule

Q: Where are the solutions publicly available?     & RQ1      \\
D: The public URL of the provided solution or study to reuse the solution.    &   \\ 
R: This question seeks the specific URLs or platforms where researchers have made their solutions or study results available. These resources allow for a practical evaluation of how easily other researchers and practitioners can access, test, and potentially replicate the study's findings. It also contributes to understanding the commitment of software engineering for the Jupyter Notebook research community to open science and sharing knowledge and tools. &   \\  \midrule
Q: Which specific software engineering problems are addressed by the study?      & RQ2     \\
D: The software engineering problem(s) that the study targets (e.g., \textit{code quality}, \textit{reproducibility} or \textit{usability}).  &   \\ 
R: This question helps to understand which challenges are focused on by software engineering researchers for Jupyter notebooks.  & \\
\midrule

Q: How do software engineering researchers evaluate their solutions?      & RQ2    \\
D: The steps to evaluate or measure the provided solution or study. &   \\ 
R: Determining how software engineering researchers evaluate their solutions in studies on Jupyter notebooks is crucial for assessing the validity and effectiveness of their findings. This question aims to understand the methodologies and metrics used for evaluation, such as experimental designs, case studies, user surveys, performance metrics, or qualitative analysis. Understanding these evaluation methods allows for critically assessing the research's reliability and applicability.  & \\
\midrule
Q: What do the authors mention as the main contributions?      & RQ2   \\
D: List of the contributions of the study.   &   \\ 
R: This question helps to map out the progress that has been made in the field so far. & \\
\midrule
Q: What do the authors mention as the implications?     & RQ2    \\
D: Implications of the results reported in the study.    &   \\ 
R: The implications are critical for understanding the real-world impact and broader significance of a study. This question seeks to uncover how the results of each study might influence future research, industry practices, educational methodologies, or software development processes. & \\
\midrule
Q: What are the limitations of the study? & RQ2      \\
D: List of the limitations presented in the study.   &   \\ 
R: This question aims to uncover the acknowledged weaknesses, constraints, or aspects that were not covered in the studies. Knowing these limitations helps evaluate the robustness of the software engineering research on notebooks. It also provides insights into areas that need further investigation or improvement in future studies. &   \\
\bottomrule 

\end{longtable}
}


\subsubsection{Developing the Review Protocol}
\label{sec:sub:reviewProtocol}
A review protocol is necessary to outline the procedure for conducting the systematic review to limit the likelihood of researcher bias~\citep{kitchenham2004procedures}. To meet these objectives, our protocol involves: (1)~defining research questions, (2)~searching and selecting primary studies, (3)~extracting data from primary studies, and (4)~synthesizing the data.

As a starting point, we formulated research questions that guided this systematic literature review. Our goal is to give an overview of the academic software engineering research on Jupyter Notebook. To achieve this, we defined two research questions (RQs):

\begin{itemize}
    \item RQ1: \textit{\rqOne} (Section~\ref{sec:slr:res:rq1})
    
    \textit{Motivation:} The first research question gives a quantitative overview of the software engineering research on Jupyter Notebook. RQ1 aims to provide a high-level overview of how the field of software engineering research on Jupyter Notebook is evolving and where it is getting attention. RQ1 also explores what kinds of solutions are being proposed by software engineering researchers to deal with problems in Jupyter notebooks. By answering RQ1, we can identify trends and patterns of software engineering research targeted at Jupyter notebooks over time. Furthermore, this question explores whether research in this area follows best practices for open science~\citep{Mendez_2020}.

    \item RQ2: \textit{\rqTwo} (Section~\ref{sec:slr:res:rq2})
    
    \textit{Motivation:} The second research question is motivated by the idea that gaining a detailed understanding of the topics explored in software engineering research on Jupyter Notebook will help to pinpoint areas that have not yet received sufficient attention and form promising future research directions. 
        
\end{itemize}

The specific procedures for searching and selecting primary studies, extracting data from primary studies, and synthesizing the data identified from these studies are presented in the next section.

\subsection{Conducting the Review}

\subsubsection{Searching and Selecting Primary Studies}\label{sec:sub:method_search}

We searched all the indexed articles in IEEE Xplore, ACM Digital Library, Elsevier ScienceDirect, Springer Nature Link, and DBLP. Together, these five sources comprehensively cover conference papers, journal articles, and other academic reports related to computer science and software engineering. 
The focus of our review is software engineering research on Jupyter Notebook. This type of research includes work that analyzes or improves current software engineering practices in notebooks. To find relevant studies for our review, we searched the title of the papers with the following query: \textit{``Notebook''} or \textit{``Jupyter''}. The literature list was compiled up to September 2025, capturing the latest software engineering research on Jupyter notebooks.


To make sure that we include only studies on software engineering research on Jupyter Notebook, we define the following inclusion criteria:  

\begin{itemize}
    \item The subject of the study should be Jupyter Notebook (and not, e.g., physical notebooks)
    \item The study should focus on at least one software engineering topic 
    \item The study must be available online to ensure its accessibility
    \item If a study has both an official and pre-print available, we picked the official one as the most recent 
    \item The study must be written in English
\end{itemize}

\subsubsection{Extracting Data from Primary Studies}\label{sec:sub:method_extract}

We created a set of questions for the data extraction form. These questions were designed to gather the necessary information from the primary studies. To streamline the data extraction process, we linked each question to one of our research questions and provided the rationale for the question. We refined this form through several iterations with randomly selected studies. Table~\ref{tab:question} presents the list of questions, their descriptions, rationales, and corresponding research questions.  
We combined the data collected from the data extraction forms based on the questions. Then, we synthesized them to answer our research questions in this SLR. This compilation of data will give an overview of the existing software engineering research on Jupyter notebooks.

\subsubsection{Synthesizing the Data}\label{sec:sub:method_synthesize}
We manually explored the title of the online version of all those papers. In this stage, we excluded several studies that did not focus on computational notebooks but on physical notebooks (e.g., laptops). 
Then, we explored the studies' abstracts and selected the papers that target software engineering-related topics on Jupyter Notebook. We also excluded studies that focused on topics that are adjacent to software engineering research on Jupyter Notebook, such as those on using Jupyter Notebook for software engineering education. For example, we excluded the experience report by \citet{al2022jupyter} on using Jupyter Notebook in classroom programming. 

We manually reviewed the list of studies to exclude duplicates, e.g., sometimes a study's final official version and unofficial pre-print version are available online. For example, Wang et al. published a pre-print of their paper titled ``Themisto: Towards Automated Documentation Generation in Computational Notebooks''~\citep{wang2021themisto}. After that, the same paper was published officially in the next year with a different title ``Documentation Matters: Human-Centered AI System to Assist Data Science Code Documentation in Computational Notebooks''~\citep{wang2022documentation}. We manually checked both versions of that paper and excluded the earlier version. 

To ensure the reliability of our study selection procedure, the first and third authors independently went through the list of studies following the procedure above. We used Cohen's Kappa statistic to measure the Inter-Rater Reliability (IRR)~\citep{mchugh2012interrater}, which indicates the level of agreement between two raters in a classification task. We measured the IRR into two stages. In the first stage, the first and third authors discussed and resolved each disagreement to come up with a list of selected studies up to 2023. In the second stage, studies after 2023 and till September 2025 were discussed and resolved by the first and second authors. The disagreements mostly arose from the first author's misunderstanding of defining the scope of provenance in the notebook. The Cohen's Kappa value was 0.73 for the first stage and 0.97 for the second stage, indicating substantial agreement in both stages. After resolving all disagreements, we identified 195 papers. We then conducted backward and forward snowballing, identifying four additional studies. Two of these had been unintentionally missed during the labeling process, which demonstrates the overall robustness of our selection procedure, as only two studies were overlooked. The other two were newly published after our cutoff date and were not yet indexed in the source databases. In total, we finalized 199 studies as the primary set for this SLR.

\subsection{Reporting the Review}
After conducting the review, we began reporting the results. Using the data extraction strategy outlined in Section~\ref{sec:sub:reviewProtocol}, we examined the primary studies to address the research questions of this literature review. We followed the data extraction questions presented in Table~\ref{tab:question} to gather relevant data for reporting that would help us answer our research questions.
We reported a quantitative overview of the software engineering research on Jupyter Notebook in Section~\ref{sec:slr:res:rq1} to address RQ1. We reported the publication year, publication type (e.g., journal, conference, symposium, workshop, and others), and industry collaborations. Our analysis also focused on studies that provided URLs for notebook replication, which allowed us to assess the hosting platforms used for these packages. Then in  Section~\ref{sec:slr:res:rq2}, we reported a comprehensive discussion of the various software engineering topics addressed in the literature we reviewed, aligning our analysis with RQ2. We searched, reviewed, and finalized a well-structured categorization of software engineering topics and subtopics. The first and third authors collaborated closely in a card-sorting-like process to categorize the studies based on different software engineering topics. We employed a Trello board\footnote{\url{https://trello.com/}} as a dynamic organizational tool to effectively manage and categorize the software engineering topics and their corresponding subtopics.

\section{RQ1: \rqOne}
\label{sec:slr:res:rq1}

\subsection{Year-wise Distributions}
\begin{figure*}[ht]
\centering
\includegraphics[width=0.8\textwidth]{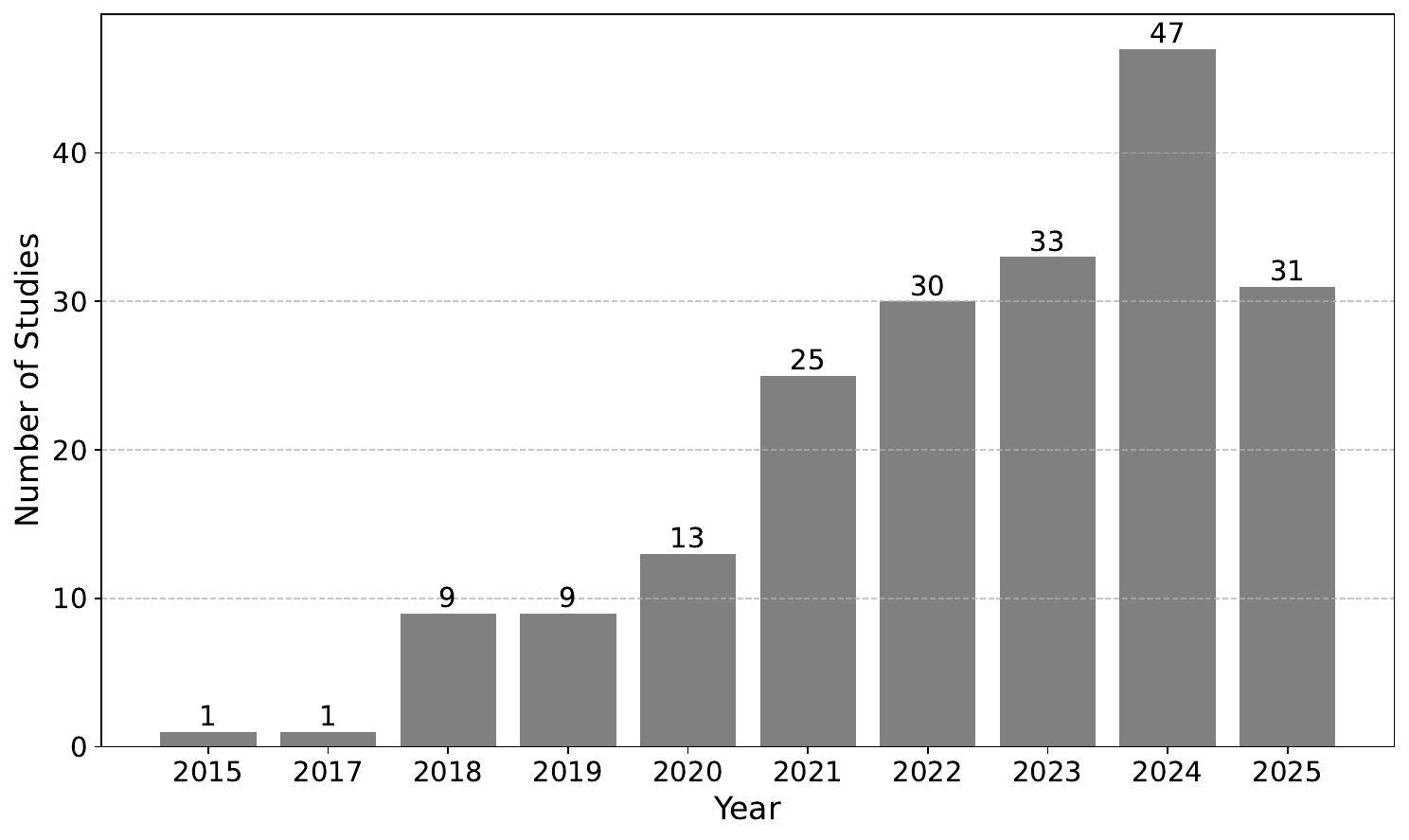}
\caption{Number of studies published over the years}
\label{fig:year}
\end{figure*}

\textbf{The data from the SLR showcases a marked increase in software engineering studies on Jupyter notebooks.} The graph in Figure~\ref{fig:year} shows the number of publications with software engineering research on Jupyter notebooks per year. The figure outlines a growing trend. Software engineering research on notebooks began with just one study in 2015 and followed an increasing trend to 47 in 2024 and 31 in 2025 (until September 25th). This trend reflects a growth in academic interest in considering software engineering practices in Jupyter notebooks. 

\begin{figure}[ht]
\centering
\includegraphics[width=0.7\textwidth]{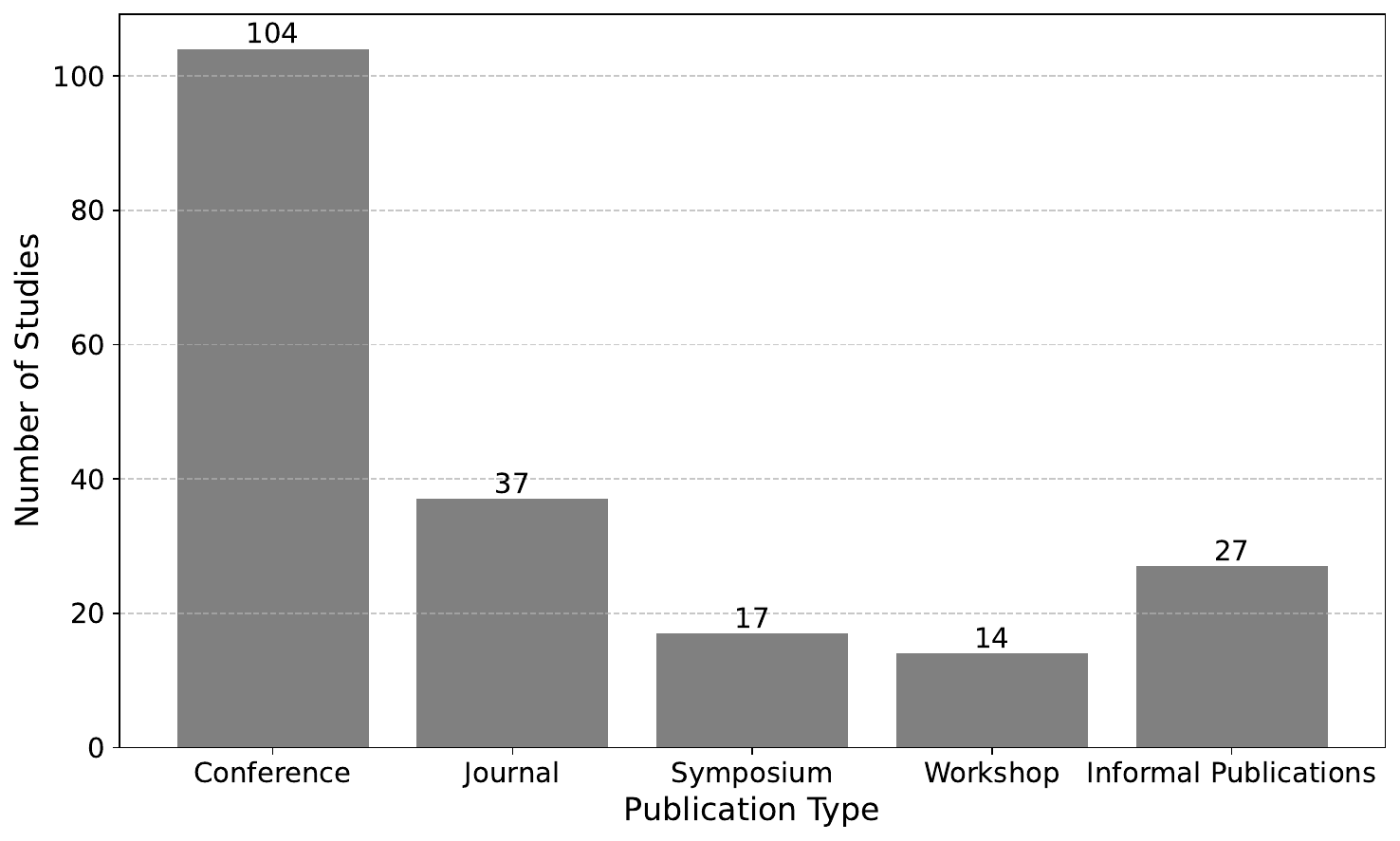}
\caption{The distribution across publication types}
\label{fig:type}
\end{figure}
\subsection{Publication Types}
\textbf{The majority of the work on software engineering research on notebooks was published at conferences.} Figure~\ref{fig:type} shows the number of studies per type. Our research findings showed that the majority, specifically 52.3\% (104 out of 199 studies), were presented at conferences. In addition, 37 studies were published in journals, 17 in symposia, and 14 in workshops. Furthermore, we identified 27 studies that appeared in informal and other types of publications, such as book chapters, thesis dissertations, or those published in the arXiv repository. It is important to note that conferences publish findings more quickly than journals, allowing researchers to share updates with the community faster~\cite{gonzalez2011articles}. The substantial number of conference papers included in the systematic literature review indicates that software engineering research on notebooks is an active research area of study.

\subsection{Publication Venues}
\textbf{The most popular venues for software engineering research on notebooks are focused on human-computer interaction (HCI)}. Table~\ref{tab:journalSymnWorkshopOthers} presents the list of conferences, journals, symposiums, workshops, and other publication venues where the primary studies have been published. We found that the most popular venue (20 studies) was the ACM Conference on Human Factors in Computing Systems (CHI). Likewise, the most popular symposium (eight studies) is the IEEE Symposium on Visual Languages and Human-Centric Computing (VL/HCC), and one of the popular journals (three studies) is the Springer International Journal on Very Large Data Bases (VLDB Journal).
This pattern emphasizes the interdisciplinary nature of working with Jupyter notebooks and underlines the importance of their user experience and usability (which are important topics at the HCI venues above).

{
\scriptsize
\setlength\LTleft{0pt} 
\setlength\LTright{0pt}
\begin{longtable}{p{1.5cm}p{1.8cm}p{5cm}p{2.3cm}r}
\caption{Publication outlets for software engineering research on notebooks in conferences, journals, symposiums, workshops and other platforms} \label{tab:journalSymnWorkshopOthers} \\
\toprule
\textbf{Type} & \textbf{Acronym} & \textbf{Venue} & \textbf{Reference} & \textbf{Count} \\
\midrule
\endfirsthead

\toprule
\textbf{Type} & \textbf{Acronym} & \textbf{Venue} & \textbf{Reference} & \textbf{Count} \\
\midrule
\endhead

\bottomrule
\multicolumn{5}{r}{\textit{Continued on next page}} \\
\endfoot

\endlastfoot
\multirow{50}{*}{Conference} &
  CHI &
  ACM Conference on Human Factors in Computing Systems &
  \cite{head2019managing, kang2021toonnote, li2021nbsearch, li2023notable,   mcnutt2023design, raghunandan2022code, rule2018exploration, wang2021makes,  wang2022stickyland, weinman2021fork, chattopadhyay2020whats, harrison2024jupyterlab, in2024evaluating, wang2024outlinespark, wang2023supernova, fang2025enhancing, fang2025large, lin2025interlink, rawn2025pagebreaks, kery2018story} &
  20 \\
   &
  ASE &
  Automated Software Engineering Conference &
  \cite{bavishi2021vizsmith, dong2021splitting, mondal2023cell2doc,   wang2020assessing, yang2022data, zhu2021restoring, huang2024code, mostafavi2024can} &
  8 \\
   &
  ICSE &
  International Conference on Software Engineering &
  \cite{patra2022nalin, quaranta2022assessing, subotic2022static, wang2021restoring, wang2020restoring, wang2020better, huang2025scientists} &
  7 \\

 &
  MSR &
  Mining Software Repositories   Conference &
  \cite{grotov2022large, quaranta2021kgtorrent, pimentel2019large, mostafavi2024distilkaggle, nakamaru2025jupyter, nguyen2025majority} &
  6 \\
   &
   TaPP &
  USENIX Conference on Theory and Practice of Provenance. &
  \cite{dao2022runtime, petricek2018wrattler, koop2017dataflow, pimentel2015collecting} &
  4 \\
  &
    e-Science &
  IEEE International Conference on eScience &
  \cite{ahmad2022reproducible, cunha2021context, li2022context, li2023dense} &
  4 \\
   &
  PEARC &
  Practice and Experience in Advanced Research Computing Conference &
  \cite{wannipurage2022framework, jayawardana2024enhancing, kim2024japper} &
  3 \\
   &
  ICPC &
  IEEE/ACM International Conference on Program Comprehension &
  \cite{robinson2022error, husak2023slicito, wong2025method} &
  3 \\
   &
  APSEC &
  Asia-Pacific Software Engineering Conference &
  \cite{ritta2022reusing, settewong2022visualize, studtmann2023histree} &
  3 \\
   &
  FSE &
  ACM International Conference on the Foundations of Software Engineering &
  \cite{dong2021qualitative, wang2024using, jin2025learning} &
  3 \\
   &
   SciPy &
  Scientific Computing with Python Conference &
  \cite{michael2022keeping, lu2020securing} &
  2 \\
 &
  SANER &
  Int. Conference on Software Analysis, Evolution and Reengineering &
  \cite{titov2022resplit, venkatesh2023enhancing} &
  2 \\
 &
   IUI &
  Annual Conference on Intelligent User Interfaces &
  \cite{muller2021data, george2024notebookgpt} &
  2 \\
 &
   ICSME &
  Int. Conference on Software Maintenance and Evolution &
  \cite{jiang2022elevating, zou2024mlpipeline} &
  2 \\
 &
  EDBT &
  International Conference on Extending Database Technology &
  \cite{horiuchi2022jupysim, watson2019pysnippet} &
  2 \\
 
 &
  CIDR &
  Int. Conference on Innovative Data Systems Research &
  \cite{macke2021automating, brachmann2020your} &
  2 \\
   &
  CAIN &
  Int. Conference on AI Engineering – Software Engineering for AI &
  \cite{quaranta2022pynblint, mostafavi2024beyond} &
  2 \\
   &
  ICMD &
  International Conference on Management of Data &
  \cite{li2024demonstration, rehman2019towards} &
  2 \\   
 &
  WWW &
  ACM Web Conference &
  \cite{kwon2023weedle} &
  1 \\
   &
  VL/HCC &
  IEEE Symposium on Visual Languages and Human-Centric Computing &
  \cite{harden2022exploring} &
  1 \\
     &
  VDS &
  IEEE Visualization in Data Science (VDS) &
  \cite{wenskovitch2019albireo} &
  1 \\
   &
  UCC &
  International Conference on Utility and Cloud Computing &
  \cite{xin2022provenance} &
  1 \\
 &
  SPLC &
  Int. Systems and Software Product Line Conference &
  \cite{brault2023taming} &
  1 \\
\multirow{44}{*}{Conference}   &
  SPLASH &
  ACM SIGPLAN conference on Systems, Programming, Languages, and Applications: Software for Humanity &
  \cite{christophe2023moon} &
  1 \\
     &
  ASSET &
  International ACM SIGACCESS Conference on Computers and Accessibility &
  \cite{christophe2023moon} &
  1 \\
    &
  SEAA &
  Euromicro Conference on Software Engineering and Advanced Applications &
  \cite{golendukhina2024unveiling} &
  1 \\
 &
   SCAM &
  IEEE Working Conference on Source Code Analysis and Manipulation &
  \cite{siddik2023code} &
  1 \\
  & 
  QCE   &
  International Conference on Quantum Computing and Engineering   &
  \cite{kinanen2024improving}   &
  1 \\
     &
  Programming &
  ACM Conference on the Art, Science, and Engineering of Programming &
  \cite{niephaus2019polyjus} &
  1 \\
     &
  PASC &
  ACM Platform for Advanced Scientific Computing Conference &
  \cite{tsai2024libyt} &
  1 \\
  &
  KDIR &
  Int. Conference on Knowledge Discovery, Engineering, and Management &
  \cite{kerzel2021towards} &
  1 \\
     &
  JCSSE &
  Int. Joint Conference on Computer Science and Software Engineering &
  \cite{ragkhitwetsagul2024typhon} &
  1 \\  
   &
   ISWC &
  International Semantic Web Conference &
  \cite{samuel2018provbook} &
  1 \\
 &
  IJCAI &
  International Joint Conference on Artificial Intelligence &
  \cite{wang2021graph} &
  1 \\
     &
  ICIIBMS &
  IEEE International Conference on Intelligent Informatics and Biomedical Sciences&
  \cite{rehman2019towards} &
  1 \\ 
  &
  HCI   &
  International Conference on Human-Computer Interaction    &
  \citep{schmieg2025enhancing}  &
  1 \\
   &
  FLAIRS &
  Int. Florida Artificial Intelligence Research Conference &
  \cite{oli2021automated} &
  1 \\
  &
  EMNLP &
  Int. Conference Empirical Methods in Natural Language Processing &
  \cite{liu2021haconvgnn} &
  1 \\
  &
  EDUCON    &
  IEEE Education Engineering    &
  \citep{bajaj2025tool}     &
  1 \\
 &
EASE &
  Int. Conference on Evaluation and Assessment in Software Engineering &
  \cite{agrawal2022understanding} &
  1 \\
 &
  CSR &
  International Conference on Cyber Security and Resilience &
  \cite{ramsingh2024understanding} &
  1 \\      
     &
  CSCL &
  International Conference on Computer-Supported Collaborative Learning &
  \cite{cai2025jupyterNotebook} &
  1 \\      

  &
  Coil Calling &
  ACM Symposium on Eye Tracking Research and Applications &
  \citep{sparmann2023jugaze} &
  1     \\
  &
  AIMLS &
  International Conference on AI ML Systems &
  \cite{werner2021bridging} &
  1 \\      
&
  ACL &
  Annual Meeting of the Association for Computational Linguistics &
  \cite{yin2023natural} &
  1 \\

  \midrule

\multirow{6}{*}{Journal} &
  EMSE &
  Empirical Software Engineering Journal &
  \cite{pimentel2021understanding, ramasamy2023visualising, venkatesh2024static, ghahfarokhi2024predicting} &
  4 \\
  &
  VLDB &
  International Journal on Very Large Data Bases &
  \cite{li2023elasticnotebook, shankar2022bolt, zhang2019juneau} &
  3 \\
   &
  TOSEM &
  ACM Transactions on Software Engineering and Methodology &
  \cite{liu2023refactoring, de2024bug, wang2024why} &
  3 \\
\multirow{34}{*}{Journal}  &
 IEEE TVCG  &
 IEEE Transactions on Visualization and Computer Graphics   &
 \cite{lin2023inksight, scully2024design, wootton2024charting}   &
 3   \\
 &
  TOCHI &
  ACM Transactions on Computer-Human Interaction &
  \cite{quaranta2022eliciting, wang2022documentation} &
  2 \\
 &
  ACM-HCI &
  Proceedings of the ACM on Human-Computer Interaction &
  \cite{rule2018aiding, weber2024extending} &
  2 \\

 &
   SoftwareX &
  SoftwareX &
  \cite{fitzpatrick2022davos, quaranta2024pynblint} &
  2 \\
 &
   CGF   &
  Wiley Computer Graphics Forum   &
  \cite{gadhave2024persist, nylund2025matplotalt} &
  2 \\
  
 &
    ACM PL &
    Proceedings of the ACM on Programming Languages &
    \cite{sato2024multiverse}   &
    1   \\
 &
  Sigmod Rec. &
  ACM SIGMOD Record Journal &
  \cite{psallidas2022data} &
  1 \\
  &

  Interactions &
  ACM Interactions &
  \cite{raghunandan2023measuring} &
  1 \\
 &
  GigaScience &
  GigaScience &
  \cite{samuel2022computational} &
  1 \\
 &
  BTW &
  Journal on Business, Technologie und Web &
  \cite{kerzel2023mlprovlab} &
  1 \\
 &
  JASEP &
  The Journal on the Art, Science, and Engineering of Programming &
  \cite{kallen2021jupyter} &
  1 \\
 &
  PLoS CB. &
  PLOS Computational Biology Journal &
  \cite{rule2018ten} &
  1 \\
 &
  TiiS &
  ACM Transactions on Interactive Intelligent Systems &
  \cite{li2023edassistant} &
  1 \\
  &
  SoftwareTT &
  Softwaretechnik-Trends Journal &
  \cite{speicher2019notes} &
  1 \\
 &
  JASIST &
  Journal of the Association for Information Science and Technology &
  \cite{candela2023approach} &
  1 \\      
&
  Softw. Pract. Exp. &
  Wiley Software: Practice and Experience &
  \cite{li2025search} &
  1 \\      
  &
  GIB &
  Software Engineering: Gesellschaft für Informatik, Bonn &
  \cite{mietchen2025analyzing} &
  1 \\      
  &
  JGI &
  Journal of Graphics Interface &
  \cite{harden2023there} &
  1 \\      
  &
  JCSE &
  Journal of Computing in Science and Engineering &
  \cite{ono2021interactive} &
  1 \\      
  &
  TVCG &
  IEEE Trans. Visualization and Computer Graphics &
  \cite{eckelt2024loops} &
  1 \\      
  &
  CCSC &
  Journal of Computing Sciences in Colleges &
  \cite{adams2023comparison} &
  1 \\

  \midrule
  \multirow{18}{*}{Symposium} &
  VL/HCC &
  Symposium on Visual Languages and Human-Centric Computing &
  \cite{koenzen2020code, chen2023whatsnext, subramanian2020casual, lau2020design, kery2018interactions, cheng2024biscuit, christophe2023moon, merino2022making} &
  8 \\
 &
  UIST &
  ACM Symposium on User Interface Software and Technology &
  \cite{wu2020b2, ouyang2024noteplayer} &
  2 \\
 &
  AISTA &
  ACM International Symposium on Software Testing and Analysis &
  \cite{venkatesh2021automated} &
  1 \\
 &
  Onward &
  ACM International Symposium on New Ideas, New Paradigms, and Reflections on Programming and Software &
  \cite{singer2020notes} &
  1 \\      
  &
  SPLASH-E  &
  ACM SIGPLAN International Symposium on SPLASH-E   &
  \cite{kimio2024kogi}  &
  1 \\
    &
  SIGCSE  &
  ACM Technical Symposium on Computer Science Education   &
  \cite{cai2025jupyter}  &
  1 \\
    &
  NOMS  &
  IEEE/IFIP Network Operations and Management Symposium  &
  \cite{ueno2020ssh}  &
  1 \\
      &
  NCA  &
  International Symposium on Network Computing and Applications    &
  \cite{faenza2024containerized}  &
  1 \\
    &
  LDAV  &
  IEEE Symposium on Large Data Analysis and Visualization  &
  \cite{savira2021writing}  &
  1 \\
  
  \midrule

\multirow{24}{*}{Workshop} &
  IDE   &
  ACM/IEEE Workshop on Integrated Development Environments  &
  \cite{titov2024hidden, wang2024dont}  &
  2 \\
  &
  IPAW &
  International Provenance and Annotation Workshop &
  \cite{koop2021notebook, samuel2021reproducemegit} &
  2 \\
  &
  SC Workshop   &
  Workshops of the International Conference for High Performance Computing, Networking, Storage and Analysis    &
  \cite{cao2024jupyter, werner2024jumper}   &
  2 \\
 &
  SOAP &
  ACM Int. Workshop on the State Of the Art in Program Analysis &
  \cite{negrini2023static} &
  1 \\
  &
  QuASoQ    &
  International Workshop on Quantitative Approaches to Software Quality &
  \cite{aydin2024cellrecommend} &
  1 \\
 &
  HILDA &
  Workshop on Human-In-the-Loop Data Analytics &
  \cite{brown2023facilitating} &
  1 \\
 &
  PAINT &
  Workshop on Programming Abstractions and Interactive Notations, Tools, and Environments &
  \cite{verano2022suppose} &
  1 \\
 &
  IWSC &
  International Workshop on Software Clones &
  \cite{sato2022comparing} &
  1 \\  
   &
  IPDPSW &
  International Parallel and Distributed Processing Symposium Workshops &
  \cite{oden2024integrating} &
  1 \\      
     &
  DL4Code &
    International Workshop on Deep Learning for Code  &
    \cite{grotov2025themisto} &
    1 \\
    &
  IDE &
  IEEE/ACM Second IDE Workshop &
  \cite{grotov2025evolving} &
  1 \\      
  
  \midrule

\multirow{9}{*}{\shortstack{Informal \\and Other\\ Publications}} &
  arXiv &
  An archive for electronic preprints of scientific studies &
  \cite{carver2025notebookos, chattopadhyay2023make, choetkiertikul2023mining, duan2023jup2kub, fangohr2020testing, grotov2024debug, grotov2024untangling, guo2024explainability, horiuchi2022similarity, in2025exploring, jiang2025exploring, jin2025suggesting, li2024kishu, li2024unlocking, li2025kernel, perez2024flexible, schroder2019reproducible, shome2024understanding, tang2025characterising, tian2025noteflow, titov2025observing, wandel2025pyevalai, weber2024computational, yang2023code, you2025datawiseagent, zhu2024facilitating} &
  26   \\
   &
  Thesis &
  Thesis dissertation was published at the University of California, San Diego &
  \cite{singer2020notes} &
  1 \\   
  \bottomrule

\end{longtable}
}


\subsection{Industry Collaborations}
\textbf{Software engineering research on Jupyter notebooks is regularly done in collaboration with industry.} Of the 199 primary studies on software engineering research on Jupyter notebooks, 37 (18.8\%) were done with industry collaborations. Table~\ref{tab:industry} presents the industry affiliations of the studies included in our SLR. Microsoft Research has the highest representation among the identified industry collaborations, with a study count of 13. JetBrains and IBM Research have seven and six studies on software engineering research on notebooks, respectively, followed by Microsoft. 
We analyzed those studies and found that Microsoft Research primarily focused on code management in notebooks~\cite{head2019managing,subotic2022static,psallidas2022data,mondal2023cell2doc,li2023notable,mcnutt2023design,lin2025interlink}. In contrast, IBM Research concentrated on improving code documentation and comprehension~\cite{liu2021haconvgnn,muller2021data,wang2022documentation,wang2021graph,wang2021makes}. JetBrains Research mainly explored tool development aimed at enhancing the execution environment and managing notebook code~\citep{titov2024hidden,titov2025observing,grotov2025themisto,titov2022resplit}. Overall, the industry affiliations reflect the widespread interest and impact of software engineering research on Jupyter notebooks in both academia and industry.

\begin{table}[t]
\footnotesize
\centering
\caption{Industry-affiliated software engineering research on Jupyter Notebook}
\label{tab:industry}
\begin{tabular}{p{4.3cm}p{8.5cm}}
\toprule
\textbf{Industry Name}	& \textbf{List of Literature} \\ 
\midrule
Microsoft Research	&  \citep{subotic2022static, psallidas2022data, head2019managing, mondal2023cell2doc, li2023notable, mcnutt2023design, weinman2021fork, chattopadhyay2020whats, chattopadhyay2023make, huang2024code, li2024unlocking, ouyang2024noteplayer, lin2025interlink} \\
JetBrains Research	&  \cite{grotov2022large, titov2022resplit, grotov2024debug, grotov2024untangling, titov2024hidden, titov2025observing, grotov2025themisto} \\
IBM Research	&  \citep{liu2021haconvgnn, muller2021data, wang2021graph, wang2021makes, wang2022documentation, cunha2021context} \\
Fujitsu Research of America	& \cite{bavishi2021vizsmith, zhu2021restoring, ueno2020ssh} \\
Google Inc	& \citep{petricek2018wrattler, yin2023natural} \\
Adobe Research  &   \citep{chen2023whatsnext, carver2025notebookos}    \\
Apple   &   \citep{cheng2024biscuit}     \\
LinkedIn    &   \citep{li2025kernel} \\
Megagon Labs    &   \citep{kwon2023weedle}   \\
Ploomber	&  \citep{michael2022keeping} \\ \bottomrule
\end{tabular}
\end{table}

\subsection{Replication Packages}
\textbf{Our systematic literature review discovered a gap in the availability of the resources necessary to replicate the research}. Specifically, more than half of the studies (112 out of 199) did not provide publicly accessible source code or installer files of their work. This made it difficult to validate and build on their findings. In contrast, the remaining 87 studies provide source code, installer files, datasets, or a combination thereof to replicate their research results. Upon a closer manual examination of these studies, we found that replication packages for 82 studies are available because they provide the source code with clear instructions to execute. The remaining five studies lacked available replication packages due to the absence of source code or lack of clear execution instructions~\cite{shankar2022bolt,ritta2022reusing, settewong2022visualize, grotov2025evolving, sparmann2023jugaze}. For example, \citet{settewong2022visualize} shared a GitHub URL\footnote{\url{https://github.com/NAIST-SE/VizJupyterNotebooks}} of their work; however, upon examination, we found neither source code nor installation files. Similarly, no code was found at another GitHub URL\footnote{\url{https://github.com/soespa/jugaze}} \citep{sparmann2023jugaze}.  

\textbf{Sharing replication packages of software engineering research on notebooks regularly does not follow open science principles.} We found that the dominant platform of choice was GitHub, with 69 studies opting to use it to host their replication packages. Although GitHub is a widely accessible platform, any changes or deletions by repository owners can be a risk to the future accessibility of the replication packages. Best practices for open science in software engineering~\cite{Mendez_2020} recommend to adopt platforms and repositories designed for the long-term archiving of research outputs. We found that only 17 studies hosted their replication packages on platforms that align with these best practices (i.e., they ensure the permanent archiving of digital artifacts). Of these, 14 were hosted on Zenodo~\citep{oli2021automated, ramasamy2023visualising, pimentel2019large, koenzen2020code, quaranta2021kgtorrent, quaranta2022eliciting, aydin2024cellrecommend, golendukhina2024unveiling, mostafavi2024distilkaggle, de2024bug, wang2024why, oden2024integrating, ono2021interactive, grotov2025themisto}, two on Figshare~\citep{liu2023refactoring, shome2024understanding} and one on OSF Preprints~\citep{eckelt2024loops}. The rest has been hosted in a commercial URL\footnote{\url{https://www.kave.cc/}}~\citep{nakamaru2025jupyter}. 

\subsection{Jupyter Notebook Extensions}
\textbf{We found that a modest 18.1\% of the studies (36 of 199) explored the development and use of Jupyter Notebook extensions to enhance the user experience within notebook environments.} Despite showing the potential benefits of Jupyter Notebook extensions, 16 studies did not provide available replication packages due to missing links or code~\cite{cunha2021context, koop2021notebook, muller2021data, watson2019pysnippet, speicher2019notes, verano2022suppose, kwon2023weedle, li2023edassistant, li2023notable, li2022context, fang2025enhancing, jayawardana2024enhancing, kim2024japper, titov2025observing, ueno2020ssh, harden2022exploring}. We found that 20 extensions were reusable and integrable into Jupyter notebooks, showcasing a variety of functionalities designed to improve the user experience of Jupyter notebooks~\cite{wang2022stickyland, bavishi2021vizsmith, kang2021toonnote, pimentel2021understanding, head2019managing, kerzel2023mlprovlab, ramasamy2023visualising, li2023elasticnotebook, perez2024flexible, gadhave2024persist, harrison2024jupyterlab, harden2023there, oden2024integrating, sparmann2023jugaze, cai2025jupyter, nakamaru2025jupyter, tsai2024libyt, eckelt2024loops, rawn2025pagebreaks, husak2023slicito}. After replicating and integrating these extensions into Jupyter notebooks, we found that the majority of Jupyter Notebook extensions are designed to enhance the visualization and presentation of notebook cells and their outputs~\cite{wang2022stickyland, bavishi2021vizsmith, ramasamy2023visualising, kang2021toonnote, head2019managing, perez2024flexible, harden2023there, eckelt2024loops, cai2025jupyter, rawn2025pagebreaks}. However, only a few studies tackle more complex challenges related to the core functionalities of Jupyter notebooks, for example, managing data provenance through their extension~\cite{gadhave2024persist, kerzel2023mlprovlab}, allowing live migration of notebook environments~\cite{li2023elasticnotebook}, and notifying potential fairness in data science code~\cite{harrison2024jupyterlab}.

\section{RQ2: \rqTwo}
\label{sec:slr:res:rq2}
We have classified studies into 11 groups based on the specific software engineering topics they addressed. By grouping those studies, we aim to uncover the key challenges and solutions in software engineering that researchers have explored in the context of Jupyter Notebook. We have listed the identified topics, subtopics, and their corresponding studies in Table~\ref{tab:SEtopics}. In this section, we will provide detailed information about these software engineering topics and their associated studies.

\begin{table*}[!tb]
\scriptsize
\caption{List of software engineering topics addressed by SE research on Jupyter Notebook}
\label{tab:SEtopics}
\begin{tabular}{p{3.3cm}p{4cm}p{5.3cm}} 
\toprule
\textbf{Topics}        & \textbf{Subtopics}                                & \textbf{List of studies} \\ \midrule




\multirow{6}{3.5cm}{Code reuse and provenance \\ (Section~\ref{sec:rq2_codereuse})}  & Code cloning                                                                  &\cite{koenzen2020code, yang2023code, kallen2021jupyter, ritta2022reusing}                         \\
                    & Reusing code snippets by code search              &\cite{horiuchi2022jupysim, li2021nbsearch, watson2019pysnippet, li2023edassistant, bavishi2021vizsmith, horiuchi2022similarity, aydin2024cellrecommend, ragkhitwetsagul2024typhon, li2024unlocking, li2023dense, bajaj2025tool, li2022context, li2025search}  \\
                    & Reproducibility                                   &\cite{pimentel2021understanding, pimentel2019large, schroder2019reproducible, rule2018ten, wang2020assessing, wang2020restoring, samuel2022computational, ahmad2022reproducible, samuel2021reproducemegit, mietchen2025analyzing, nguyen2025majority}                          \\
                    & Provenance                             &\cite{koop2021notebook, pimentel2015collecting, dao2022runtime, kerzel2023mlprovlab, kerzel2021towards, samuel2018provbook, kery2018interactions, gadhave2024persist, harrison2024jupyterlab, studtmann2023histree, fang2025enhancing, eckelt2024loops, kery2018story}              \\ \midrule
\multirow{5}{3.5cm}{Managing computational environment and workflow (Section~\ref{sec:compWorkflow})}     & Empirical studies on workflows                                   &\cite{lau2020design, raghunandan2023measuring, choetkiertikul2023mining, psallidas2022data, rehman2019towards, koop2017dataflow, subramanian2020casual, golendukhina2024unveiling, zou2024mlpipeline, sparmann2023jugaze}  \\
                    & Computational environment in notebooks                             &\cite{duan2023jup2kub, cunha2021context, chattopadhyay2020whats, lu2020securing, li2023elasticnotebook, weinman2021fork, ramsingh2024understanding, kinanen2024improving, cao2024jupyter, titov2024hidden, li2024demonstration, zhu2024facilitating, sato2024multiverse, li2024kishu, jayawardana2024enhancing, kim2024japper, zhang2019juneau, tsai2024libyt, christophe2023moon, carver2025notebookos, ueno2020ssh, savira2021writing}                          \\
                    & Managing dependencies                     &\cite{zhu2021restoring, wang2021restoring, fitzpatrick2022davos, wannipurage2022framework}                          \\
                    & Performance analysis                              &\cite{werner2021bridging, werner2024jumper, faenza2024containerized, oden2024integrating, prathanrat2018performance, grotov2025themisto}                          \\ \midrule
\multirow{3}{3.5cm}{Readability of notebooks (Section~\ref{sec:rq2_readability})}   & Refactoring                                                                                                   &\cite{liu2023refactoring, sato2022comparing, dong2021qualitative, dong2021splitting, titov2022resplit, shankar2022bolt, head2019managing}                          \\
                    & Non-linear visualization                          &\cite{wang2022stickyland, kang2021toonnote, rule2018aiding, chattopadhyay2023make, merino2022making, rawn2025pagebreaks, husak2023slicito, grotov2025evolving, harden2022exploring, harden2023there}                          \\  
                    & Accessibility     &   \cite{venkatesh2023notably, nylund2025matplotalt}     \\
                                            \midrule

\multirow{5}{3.5cm}{Documentation of notebooks \\(Section~\ref{sec:rq2_doc})} & Empirical studies on documentation                                                                   &\cite{rule2018design, rule2018exploration, raghunandan2022code, wong2025method}                          \\
                                            & Documentation generation                          &\cite{wang2022documentation, wang2021graph, liu2021haconvgnn, wang2021makes, mondal2023cell2doc, muller2021data, li2023notable, lin2023inksight, mostafavi2024beyond, wang2024outlinespark, ouyang2024noteplayer, mostafavi2024can, schmieg2025enhancing, lin2025interlink}              \\
                                            & Cell header generation                                 &\cite{venkatesh2023enhancing, venkatesh2021automated, venkatesh2024static, perez2024flexible}                    
                        \\  \midrule
\multirow{4}{3.5cm}{Testing and debugging (Section~\ref{sec:testing})}       & Empirical studies on testing and debugging          &\cite{de2024bug, shome2024understanding, wang2024why}                          \\
                                            & Detecting bugs                                     &\cite{robinson2022error, xin2022provenance, brachmann2020your, patra2022nalin, fangohr2020testing, wang2024using, grotov2024untangling, grotov2024debug, tang2025characterising, jiang2025exploring, titov2025observing}                          \\
                                            & Detecting data leakage                            &\cite{yang2022data, subotic2022static, negrini2023static}           \\    \midrule
\multirow{2}{4.3cm}{Visualization in notebooks (Section~\ref{sec:rq2_visual})} & Empirical studies on visualizations                                                                     &\cite{agrawal2022understanding, settewong2022visualize,                                                       ramasamy2023visualising, wang2023supernova, wootton2024charting, wenskovitch2019albireo, christophe2023moon, tian2025noteflow}               \\
                                            & Interactive visualization                           &\cite{kwon2023weedle, wu2020b2, scully2024design, in2024evaluating, guo2024explainability, cai2025jupyter, ono2021interactive}       \\ \midrule

\multirow{3}{3.5cm}{Best practices \\ (Section~\ref{sec:bestPractice})}             & Following code style standards                                                                                &\cite{grotov2022large, wang2020better, siddik2023code, candela2023approach, quaranta2022pynblint, quaranta2024pynblint, speicher2019notes, adams2023comparison, huang2025scientists, cai2025jupyterNotebook}                          \\
                                            & Best practices for collaborative use &\cite{quaranta2022assessing, quaranta2022eliciting, wang2024dont}         \\    \midrule
\begin{tabular}[c]{@{}l@{}}Cell execution order \\ (Section~\ref{sec:rq2_order})\end{tabular}                        &      \begin{tabular}[c]{@{}l@{}} - \end{tabular}                                                                                     &\cite{oli2021automated, singer2020notes, jiang2022elevating,                                                  michael2022keeping, macke2021automating, brown2023facilitating, in2025exploring} \\     \midrule                                            

\begin{tabular}[c]{@{}l@{}}AI-based coding assistance \\for notebooks \\ (Section~\ref{sec:rq2_codegen})\end{tabular}                    &   \begin{tabular}[c]{@{}l@{}}- \end{tabular}                        &\cite{yin2023natural, brault2023taming, mcnutt2023design, chen2023whatsnext, george2024notebookgpt, huang2024contextualized, cheng2024biscuit, weber2024computational, kimio2024kogi, you2025datawiseagent, fang2025large, jin2025learning, wandel2025pyevalai, jin2025suggesting}                          \\  \midrule

\begin{tabular}[c]{@{}l@{}}Supporting other \\programming paradigms \\ (Section~\ref{sec:rq2_support})\end{tabular} & -                               &\cite{niephaus2019polyjus, petricek2018wrattler, verano2022suppose, weber2024extending, li2025kernel} \\ \midrule

\multirow{3}{3.5cm}{Datasets of notebooks (Section~\ref{sec:rq2_datasets})}                      &   \multirow{3}{4.3cm}{-}                           &\cite{grotov2022large, quaranta2022eliciting, li2021nbsearch, liu2021haconvgnn, wang2021restoring, quaranta2021kgtorrent, koenzen2020code, pimentel2019large, raghunandan2022code, de2024bug, liu2023refactoring, ramasamy2023visualising, mostafavi2024distilkaggle, shome2024understanding, golendukhina2024unveiling, huang2024contextualized, grotov2024untangling, wang2024why, aydin2024cellrecommend, ghahfarokhi2024predicting, jin2025suggesting, nakamaru2025jupyter}                          \\ 
                                            \bottomrule
                                            
\end{tabular}
\end{table*}

\subsection{Code Reuse and Provenance}\label{sec:rq2_codereuse}
Code reuse in Jupyter notebooks refers to the intentional act of leveraging previously written code cells. Code reuse within Jupyter notebooks differs from traditional IDEs, particularly regarding the notebook's structure and workflow. Jupyter notebooks employ a cell-based execution model that facilitates iterative coding, making them well suited to quickly test hypotheses and iterate on data by reusing existing code cells~\cite{koenzen2020code}. A survey among Microsoft data scientists revealed that 94\% of the participants considered reusing existing code in Jupyter notebooks to be at least ``important''~\cite{chattopadhyay2020whats}. 
Users often reuse code through code cloning, where cells are copied within or across notebooks, which can lead to inconsistencies and technical debt~\cite{koenzen2020code}. To find reusable code more efficiently, code search tools provide support by helping users locate relevant code cells or workflows~\cite{li2021nbsearch}. However, proper provenance management is crucial for understanding the origin and evolution of reused code, ensuring that dependencies and execution history are clear~\cite{xin2022provenance}. Moreover, reproducibility becomes a challenge in notebooks, as reused code might not produce consistent results if dependencies or execution orders are not properly tracked~\cite{wang2020assessing}. 
In this subsection, we describe these studies that address the challenges and opportunities of reusing code within Jupyter notebooks and provenance.

\subsubsection{Code Cloning}
Code cloning in Jupyter notebooks refers to copying and pasting code cells within the same notebook or between different notebooks. Research has shown that code cloning is common in Jupyter notebooks. Over 70\% of all notebook code cells on GitHub are exact copies of other cells, and approximately 50\% of all notebooks contain no unique cells~\cite{kallen2021jupyter}. These findings are also supported by~\citet{yang2023code}, who noted that about 32\% of Jupyter notebooks hosted in GitHub repositories were directly copied from Stack Overflow.  
The code snippets that are most frequently cloned in Jupyter notebooks mainly relate to visualization (21\%), followed by ML (15\%)~\cite{koenzen2020code}. Among reused code, interproject clones in Jupyter notebooks are far more common than intraproject clones, which is the opposite of the prevalence of code clones in Java code~\cite{gharehyazie2019cross}. Furthermore, another study indicated that top-level notebook users (i.e., Grandmasters in Kaggle) are most likely to clone common abstractions such as importing packages, configurations, file IO operations, and showing data~\cite{ritta2022reusing}.

\subsubsection{Reusing Code Snippets by Code Search}\label{sec:rq2_codesearch}

Searching for code snippets in Jupyter notebooks can be different from traditional code. While users often seek specific functions or APIs in traditional code, notebook users frequently search for code cells~(snippets) that analyze similar data or follow comparable workflows~\citep{li2021nbsearch}. In this context, semantic code search is vital, which enables natural language queries that reflect the intended functionality of code cells rather than just API names or keywords~\citep{li2021nbsearch}. Moreover, the flexible nature of notebooks, where code, data, and results are integrated within a single file, contrasts with the modular organization of traditional code. This integration in notebooks makes it more difficult to isolate functionality and understand the context of individual cells, which imposes challenges for code search such as tracking the logical flow of code and capturing enough context within a code cell~\citep{head2019managing}.

Despite sharing common objectives, code search tools for Jupyter notebooks vary in their methodological approaches. For instance, some tools utilize keyword-based search mechanisms, where users input specific terms related to their queries. This approach, demonstrated in research by \citet{watson2019pysnippet}, allows users to find relevant code by matching their search terms with corresponding code repositories and snippets. On the other hand, certain advanced search tools leverage deep learning techniques to interpret natural language queries. This enables users to search for code snippets more intuitively, using conversational language instead of strict keywords~\citep{li2021nbsearch}. Similarly, \citet{ragkhitwetsagul2024typhon} proposed \textit{Typhon}, which integrates traditional information retrieval techniques~(BM25) with UniXcoder~\citep{guo2022unixcoder} code embeddings to perform text-to-code matching. In addition, \citet{li2024unlocking} developed a semantic search framework explicitly tailored for Jupyter notebooks, leveraging embeddings generated by large language models~(LLMs) from both markdown and code cells.

Additionally, some tools enhance search efficacy by employing graph representations to illustrate code flow relationships within a Jupyter notebook. By mapping the interactions and dependencies between different code cells, these graph-based tools can identify interrelated code cells and suggest connections that the user needs~\citep{horiuchi2022jupysim}. \textit{EDAssistant}~\citep{li2023edassistant} introduces a methodology for context-aware code search in Jupyter Notebook. This approach prioritizes example notebooks that align with the user's intent and existing code, assisting users in finding relevant code snippets to reuse and adapt during exploratory data analysis.
Similarly focused on cell-level recommendation, \citet{aydin2024cellrecommend} proposed a specialized cell recommendation approach targeting ML notebooks.
Furthermore, some search solutions specifically target the data visualization domain by mining, extracting, and cataloging Python functions designed for visualization purposes~\citep{bavishi2021vizsmith}. This approach leverages reusable visualization snippets from data science notebooks to facilitate access to visualization-related code, thereby promoting the reuse and adaptation of existing visualization workflows.  

Continuing this trend, new approaches have emerged to improve code search and reuse in notebooks. For instance, a system called \textit{DeCNR} was developed to fuse sparse and dense retrieval models to effectively locate relevant computational notebooks~\citep{li2023dense}. In addition, the authors introduced a dedicated evaluation dataset of notebook queries with relevance judgments to enable fair benchmarking of search techniques. Similarly, a context-aware notebook search framework was developed to help researchers seamlessly discover external notebooks relevant to their current analysis, integrating semantic search capabilities into a Jupyter-based virtual research environment~\citep{li2022context}. In educational settings, \citet{bajaj2025tool} presented \textit{JBEval}, a tool that identifies similar code and text blocks across a collection of notebooks to detect reuse or plagiarism, providing similarity scores for code and text and visualizing identical segments for further inspection. Furthermore, a unified search system was introduced that allows users to query multiple types of research assets (including code snippets, datasets, and documentation) directly from within a notebook, reflecting a move toward richer, context-integrated search functionalities in Jupyter environments~\citep{li2025search}.

\begin{table}[t]
\caption{Good and bad practices that affect the reproducibility of a notebook}
\label{tab:nbGoodBadPractice}
\scriptsize
\begin{tabular}{p{13cm}} 
    \textbf{\underline{Good practice \#1: Share and explain the data}}~\citep{rule2018ten} \\
    Reproducibility requires sharing data alongside notebooks. When full datasets are too large or sensitive to share, provide a sample or detailed descriptions of the data and processing steps. Breaking complex datasets into tiers ensures interpretability while maintaining accessibility and reproducibility. \\
    \addlinespace[0.5em]
    \textbf{\underline{Good practice \#2: Document the process, not just the results}}~\citep{rule2018ten} \\
     Notebooks require documenting the process throughout the analysis, including taking notes during the analysis, capturing key decisions, reasoning, and observations, and preserving the context of the work. This approach enhances the clarity and utility of notebooks, facilitating their use for future reference and collaboration. \\
    \addlinespace[0.5em]
    \textbf{\underline{Good practice \#3: Record and manage library dependencies}}~\citep{rule2018ten} \\
     The notebook format does not encode library dependencies with pinned versions, making it difficult (and sometimes impossible) to reproduce a notebook. To ensure reproducibility, users of notebooks should carefully manage their dependencies using an environment management package. This process involves creating and sharing a file, such as `environment.yml' or `requirements.txt', which lists all the libraries and their specific versions used in the notebook. Such a file offers a clear and accurate description of these dependencies. \\

     \addlinespace[1em]
    \textbf{\underline{Bad practice \#1: Presence of non-executed code cells}}~\citep{pimentel2021understanding} \\
     Non-executed cells can lead to discrepancies between the notebook's apparent logic and its actual state, making it unclear whether the code has been tested or contributes to the presented results. This practice hinders the ability of others to reproduce the workflow, as the notebook may fail to run as expected or produce inconsistent outcomes. \\
     
     \addlinespace[0.5em]
    \textbf{\underline{Bad practice \#2: Out-of-order cell execution}}~\citep{pimentel2021understanding} \\
     Out-of-order cell execution can create challenges for others trying to trace the steps necessary to reproduce results. This occurs because dependencies between cells may need to be clarified or temporarily broken. Such practices can lead to errors, incomplete outputs, or incorrect conclusions when the notebook is reproduced. \\

     \addlinespace[0.5em]
    \textbf{\underline{Bad practice \#3: Presence of hidden states}}~\citep{pimentel2019large} \\
     A hidden state in a notebook occurs when variables or data are changed in ways that are not clearly documented. Hidden states caused by cell re-execution or removal make the notebooks skip numbers in the execution counter sequence. This often results from executing cells out of order or from relying on previous sessions. Figure~\ref{fig:hiddenState} presents an example of a hidden state. Such hidden states can make it difficult for others to understand or replicate the workflow since the notebook's output depends on undocumented or inaccessible conditions. \\

\end{tabular}
\end{table}

\subsubsection{Reproducibility}

Reproducibility ensures that other notebook users can get the same results from a notebook by following the same steps, using the same data, and working in the same environment~\cite{ahmad2022reproducible}. Reproducibility differs from code reuse and cloning, which typically involves adapting code or directly copying code into different environments and applying the code to different data to meet different objectives of users, resulting in outputs that are different from the original notebook.
 
Reproducibility is known as one of the key promises of Jupyter notebooks, as they are often used to share results with others, allowing users to trace the steps from raw data to final results~\cite{kluyver2016jupyter}. However, achieving reproducibility remains challenging due to issues such as out-of-order execution and repetitive execution of the same cell~\cite{wang2020assessing}. These issues differ from traditional programming, where code execution is typically linear~(from top to bottom), ensuring a clear and consistent execution flow. For example, notebook users might execute the same cell multiple times and only save the latest execution state, causing cells located at the notebook's beginning to be executed after later cells. This nonsequential execution can confuse others trying to reproduce results, affecting notebook reproducibility~\cite{wang2020assessing, wang2020restoring}. 

A study by \citet{pimentel2019large} analyzed more than 1.4 million Jupyter notebooks from GitHub and found significant reproducibility issues. The study noticed that only 25\% of the valid notebooks executed without errors under straightforward conditions, and only 5\% reproduced results identical to their original outputs. Furthermore, \citet{schroder2019reproducible} revealed that only 14\% of the available notebooks in published academic research on PubMed,\footnote{\url{https://pubmed.ncbi.nlm.nih.gov/}} where Jupyter notebooks are used as implementation, were reproduced successfully. The study noticed the need for comprehensive documentation and containerization of notebooks, which supports the good practices presented by \citet{rule2018ten}. In a more recent large-scale analysis, \citet{mietchen2025analyzing} attempted to automatically rerun thousands of notebooks associated with biomedical research articles in PubMed Central, highlighting many recurring failures (such as missing dependencies and improper execution order) and noting how the corpus of notebooks evolved over time. However, not all notebooks that fail to run are beyond repair. \citet{nguyen2025majority} revisited prior claims that roughly 76\% of public notebooks are non-executable, arguing that this estimate relies on an overly rigid notion of executability. Analyzing 42,546 notebooks, they showed that only about 21\% are truly non-restorable. For the remaining notebooks, executability can be improved by 43\% through installing missing dependencies and by 28\% through generating synthetic data. This suggests that notebooks have higher latent reproducibility than previously reported, provided that tools or users address minor reproducibility pitfalls. We have analyzed the relevant studies and presented a list of good and bad practices that affect the reproducibility of a notebook in Table~\ref{tab:nbGoodBadPractice}.

Researchers have demonstrated that the reproducibility of Jupyter notebooks can be significantly improved by resolving cell dependencies within the notebook's code~\cite{wang2020assessing, wang2021restoring}. Cell dependencies represent the relationships between variables, functions, and data in different code cells. \citet{wang2020restoring} proposed a method to build a cell dependency graph using static analysis to accurately model these relationships and found that 83\% of executable notebooks can be reproduced using their approach.  \citet{samuel2021reproducemegit} developed a visualization tool \textit{ReproduceMeGit} to check the reproducibility of Jupyter notebooks from GitHub by extending the work from \citet{pimentel2019large}. This tool can automatically install dependencies from \textit{requirements.txt} or \textit{setup.py} files and execute the notebook to provide highlights of the reproducibility study through a user interface.

\begin{figure}[t]
\centering
\includegraphics[width=.8\textwidth]{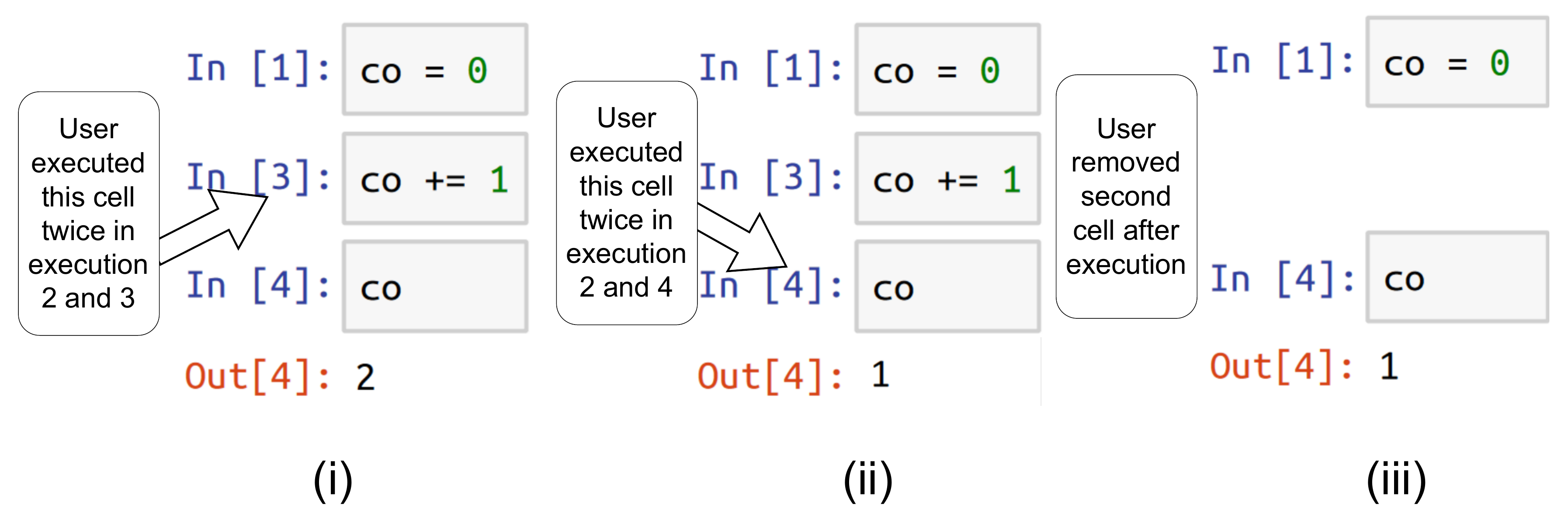}
\caption{An example of a hidden state in Jupyter Notebook with three different execution variations}
\label{fig:hiddenState}
\end{figure}

\subsubsection{Provenance}
Provenance in notebooks refers to the detailed history of actions, executions, and dependencies that led to the current state of a notebook. It includes the sequence of cell executions, the modifications made to cells, variable dependencies, data flow, and intermediate outputs~\cite{kerzel2023mlprovlab}. Unlike reproducibility which emphasizes achieving identical results under a consistent initial environment, provenance addresses the historical and iterative details present in the notebook's creation and evolution. Managing provenance in Jupyter Notebook differs from traditional systems due to its iterative and non-linear nature. In a notebook, users can execute cells in any order, creating challenges in tracking the execution history and dependencies~\citep{kerzel2021towards}. 
Empirical observations underscore this challenge, for example, \citet{kery2018story} found that data scientists often explore ideas non-linearly and later refactor their notebooks into a linear narrative (the ``story'' in the notebook), which can obscure the trial-and-error process. This highlights the need for provenance tools to capture the actual sequence of analysis steps. 
Additionally, Jupyter notebooks merge code, data, and outputs into a single file, whereas traditional systems follow a sequential execution process that generates output externally. As traditional systems offer clear execution pipelines and distinct output demonstrations, provenance management becomes more straightforward. In contrast, Jupyter notebooks require specialized tools for managing provenance to capture dynamic state changes and overwritten variables~\cite{kerzel2023mlprovlab, koop2021notebook}. 

Researchers developed provenance detection techniques specialized for Jupyter Notebook to reuse notebooks by obtaining the original results. In the early stage of Jupyter Notebook usage, \citet{pimentel2015collecting} proposed a notebook tool to overcome the limitations in provenance support by automatically capturing and analyzing the execution history and environment settings of code within notebooks. The study extends the \textit{noWorkFlow} tool~\cite{murta2015noworkflow} developed for Python scripts to capture the provenance of scripts, including control flow information and library dependencies. The proof can be inferred by statically analyzing code cells on execution counts and cell positions within the notebooks~\cite{koop2021notebook}. A scheduler that adjusts the order of cell execution based on data dependencies detected during runtime can be used to improve the accuracy of provenance of Python code in notebooks~\cite{dao2022runtime}.  

In addition to focusing on tools and execution kernels, researchers have developed extensions to capture provenance. \textit{MLProvLab}~\citep{kerzel2023mlprovlab} is a JupyterLab extension to track, manage, compare and visualize the provenance of computational experiments. It can capture the provenance at run-time by detecting dependency graphs. Another notebook extension, \textit{ProvBook}, was designed to capture and visualize the provenance of notebook executions~\cite{samuel2018provbook}. This extension automatically stores and visualizes provenance information within the notebook's metadata, detailing execution times, inputs, and outputs for each cell. It helps users to see how data and analysis have evolved over time and compare results, which helps to enable notebook reproducibility. \citet{kery2018interactions} developed an extension named \textit{Verdant} that records the history of all previous changes in a notebook and quickly retrieves versions of a specific artifact from the existing versions of the entire document. It can help users to visually compare side-by-side multiple versions of different notebook artifacts, including code cells, tables, and images, which are absent in traditional version control systems like Git. Building on the idea of visualizing notebook evolution, \citet{eckelt2024loops} introduced a system called \textit{Loops} that leverages provenance to support exploratory data analysis. Loops captures the evolution of a notebook's state as cells are edited and re-executed, and presents an interactive timeline visualization of these changes. This allows users to replay or inspect how each modification to the code or data affected the results, thereby aiding in understanding the analytical process and preserving intermediate insights.

Expanding on these efforts, researchers have explored versioning mechanisms for notebooks. \citet{studtmann2023histree} introduced \textit{Histree} for automatic versioning and visual branching of notebook modifications. It organizes experiment histories into tree-based structures for easier navigation. Similarly, \citet{fang2025enhancing} proposed a two-dimensional version control approach (realized in a tool called \textit{Kishuboard}) that captures both code and data state in notebooks. By adjusting ``code'' and ``data'' axes (or knobs), users can intuitively branch, roll back, and compare multiple divergent notebook states, achieving flexible exploration without losing past states; user studies showed that such integrated code-and-data versioning significantly enhanced productivity in complex data analysis tasks. Further extending provenance into interactive visual analytics, \citet{gadhave2024persist} proposed \textit{Persist}, which captures interaction provenance across code and visualizations. Empirical evaluations showed improvements in analysis efficiency, accuracy, and reproducibility. Finally, \textit{Retrograde}~\cite{harrison2024jupyterlab} integrates provenance tracking with fairness auditing. It provides real-time, context-aware notifications triggered by notebook events and maintains a dynamic data ancestry graph, helping users detect and address fairness and bias concerns during data preprocessing and modeling.

\begin{tcolorbox}[colback=gray!5, colframe=black, title=Summary of code reuse and provenance ]

Code reuse in Jupyter notebooks is essential for improving productivity and collaboration while addressing the challenges associated with their unique structure and iterative workflows. Users can more effectively leverage existing resources and maintain reusable notebooks through techniques such as code cloning, context-aware code search, provenance management, reproducibility practices, and advanced versioning tools. The development of innovative solutions, such as context-sensitive code search tools~\citep{li2021nbsearch, li2023edassistant}, code dependency resolution systems~\citep{wang2020restoring}, and provenance tracking extensions~\citep{kerzel2023mlprovlab, pimentel2015collecting, harrison2024jupyterlab}, highlights ongoing efforts to support effective code reuse and provenance. By integrating these solutions and best practices (see Table~\ref{tab:nbGoodBadPractice}), notebooks can become more reusable and maintainable, thus establishing a stronger foundation for data science workflows and collaborative research.
\label{summaryCodeReuse}
\end{tcolorbox}

\subsection{Managing Computational Environment and Workflow}
\label{sec:compWorkflow}
Managing the computational environment and workflow in Jupyter notebooks is crucial to ensure the successful execution and reproducibility of data science tasks~\citep{psallidas2022data, lu2020securing}. Unlike traditional programming environments, which are typically well defined and managed by integrated development environments or build systems, a Jupyter Notebook does not have a separate file to manage the configuration~\citep{kluyver2016jupyter}. Traditional programming environments commonly include a configuration file~(e.g., ``requirements.txt'' in Python) to ensure a consistent and reproducible setup for code execution. In contrast, Jupyter notebooks execute code cells independently, and each notebook has its own dependencies~\citep{cunha2021context, raghunandan2023measuring}. This section explores various aspects of computational environment management, including empirical studies on notebook workflows, the impact of different environments on code behaviour, and the challenges associated with managing library dependencies.

\subsubsection{Empirical Studies on Workflows}

\citet{lau2020design} analyzed 60 notebooks and summarized four main stages of a data science workflow: importing data into notebooks, writing and editing code, running the code to generate output, and publishing the results. However, workflows can vary. For example, notebooks often start with exploration, where users write lots of code to find interesting patterns in the data, and end with a presentation or explanation~\citep{raghunandan2023measuring}. 
In addition, recent empirical analyses have shed further light on workflow diversity. \citet{golendukhina2024unveiling} studied 138,376 Kaggle notebooks and identified significant variations in data preprocessing practices correlated with user expertise, highlighting a notable gap between model-centric activities and actual data cleaning efforts. Similarly, \citet{zou2024mlpipeline} uncovered that data scientists frequently and manually experiment with alternative ML pipeline configurations (e.g., data preparation, model selection), emphasizing the limitations of managing and systematically exploring pipeline variations in traditional notebook environments. 
On the other hand, \citet{psallidas2022data} identified two types of pipelines in Jupyter notebooks that are key components of data science workflows: explicit and implicit. Explicit pipelines utilize tools like \textit{sklearn.pipeline} to define structured steps for tasks such as data pre-processing and model training. In contrast, implicit pipelines rely on ad hoc function calls with libraries such as Pandas to clean, merge, and visualize data~\citep{psallidas2022data}. 
Researchers have also suggested ways to map workflows to better understand how notebooks work. One approach uses directed acyclic graphs to show the flow of data and tasks within a notebook~\citep{rehman2019towards}. These graphs can highlight where variables are reused (if a cycle exists) or how tasks are connected. Another solution is a Jupyter Notebook extension that assigns unique IDs to each cell, making it easier for users to track how cells depend on each other~\citep{koop2017dataflow}. These tools help manage and simplify the often complex workflows in notebooks. 
To further understand user behaviour, \citet{sparmann2023jugaze} introduced an eye-tracking and logging tool named \textit{JuGaze} which records how users visually navigate notebook cells. This fine-grained analysis of user attention offers insights into workflow patterns and can inform interface designs to better support common usage behaviours.

\subsubsection{Computational Environments in Notebooks}
The computational environment in notebooks refers to the environment in which Jupyter notebooks run~\citep{kluyver2016jupyter}. This environment can affect the code execution process because different execution environments may have different dependencies, causing inconsistencies in code behaviour~\citep{duan2023jup2kub}. Researchers have found that data scientists face computational environment-related challenges throughout the analytics workflow, from setting up the notebook to deploying it to production~\citep{chattopadhyay2020whats}.

Although notebooks typically run in a single computational environment, there are advantages to running a notebook in multiple computational environments~\citep{cunha2021context, duan2023jup2kub}. For example, parts of a notebook, such as model training algorithms, may require specialized computational resources that are unavailable in a standard environment designed primarily for data exploration or visualization. \citet{cunha2021context} presented a solution developed as a Jupyter Notebook extension to automatically determine and migrate selected notebook cells to appropriate computational environments for execution. Their solution leverages abstract syntax trees and dependency tracking to extract the selected notebook cells and the dependencies. Another study discussed the possibility of automating the migration of a computational environment for Jupyter notebooks to a distributed Kubernetes\footnote{\url{https://kubernetes.io/}} environment~\citep{duan2023jup2kub}. This approach aims to overcome the limitations of Jupyter notebooks, such as scalability and fault tolerance, by dividing notebooks into executable steps and deploying them in a \textit{Kubernetes} cluster. In addition, \textit{ElasticNotebook}~\citep{li2023elasticnotebook, li2024demonstration} provides live migration through optimized checkpointing and restoration to reduce migration overhead.
Expanding this concept further, \citet{kinanen2024improving} introduced a custom Jupyter kernel to simplify offloading quantum computation tasks directly from the notebook to remote Kubernetes clusters, thus lowering access barriers to high-performance quantum computing without requiring deep infrastructure expertise.

Moreover, bridging notebooks with high-performance computing (HPC) resources has become an area of focus. \citet{jayawardana2024enhancing} developed a gateway named \textit{Cybershuttle} that enables seamless integration of personal Jupyter notebooks with remote HPC and cloud infrastructures, allowing researchers to offload heavy computations to supercomputers while using familiar notebook interfaces. In the scientific computing domain, notebooks have been used to write and manage large-scale simulations. For instance, \citet{savira2021writing} describe strategies for running and analyzing large scientific simulations via Jupyter, and \citet{tsai2024libyt} introduced \textit{Libyt}, which integrates the \textit{yt} analysis library with Jupyter for parallel in situ analysis of simulation data. To improve GPU utilization for interactive model training in notebooks, \citet{carver2025notebookos} proposed \textit{NotebookOS}, an operating system for notebooks that replicates kernel processes across multiple GPU servers and dynamically allocates GPUs only during cell execution, significantly reducing idle GPU time and cost in cloud-based notebook services. Additionally, \citet{ueno2020ssh} developed an SSH-based Jupyter kernel for remote infrastructure administration, enabling users to execute shell commands on remote servers from within a notebook. This extends the notebook’s reach to managing external systems directly through the notebook environment.

Recognizing security as a crucial factor in computational environment design, researchers have explored specific strategies for improving Jupyter notebook security. For instance, \citet{lu2020securing} proposed an architectural security framework leveraging containerization, load balancing, authentication, and encryption mechanisms, thus safeguarding sensitive data within cloud-based notebook execution environments. In addition, \citet{ramsingh2024understanding} conducted empirical analyses revealing gaps in current notebook security practices, such as poor awareness of threats and insufficient technical expertise. Based on their findings, they proposed the Jupyter multi-layer security~(JMLS) defence model explicitly designed to reinforce notebook security. Complementing these perspectives, \citet{cao2024jupyter} developed a systematic taxonomy to classify and better understand the variety of network-based security threats impacting Jupyter notebooks, especially in high-performance computing~(HPC) scenarios. Their exhaustive analysis highlighted vulnerabilities involving ransomware, data exfiltration, misconfiguration, and resource misuse, providing insights for secure management of computational environments.

Researchers have also investigated various computational environments and techniques to tackle Jupyter Notebook reproducibility and usability issues. \citet{sato2024multiverse} addressed reproducibility issues arising from potentially unsafe dynamic notebook modifications by developing \textit{Multiverse Notebook}, an environment (initially introduced as a prototype by \citet{christophe2023moon} that enables safe and efficient ``time-travel''~(cell-wise checkpointing) based on the POSIX ``fork()'' mechanism. In this design, each executed cell runs as a separate process and the collection of processes forms a tree representing the entire notebook state. To address hidden state and out-of-order execution issues, \citet{weinman2021fork} created a notebook extension that allows for easy branching (or ``forking'') of execution states into separate kernel instances to simplify concurrent exploration of different analytical approaches. Furthermore, \citet{li2024kishu} introduced \textit{Kishu}, which employs namespace-patching techniques to efficiently revert or undo notebook states without resorting to costly kernel restarts or complete re-executions. Complementing these approaches, \citet{zhang2019juneau} presented \textit{Juneau}, which repurposes the notebook backend as a data lake management layer that 
integrating provenance-aware indexing and reuse of code and data artifacts directly into Jupyter to enhance usability.

To improve Jupyter notebook usability and integration into software development workflows, \citet{titov2024hidden} explored strategies for integrating computational notebooks within IDEs to better align interactive and exploratory notebook usage with standard software engineering practices. Similarly, \citet{kim2024japper} presented \textit{Japper}, a framework that streamlines the conversion of Jupyter notebooks into standalone scientific web applications, thereby simplifying the transition from ad hoc analyses to deployable, interactive tools. In addition, acknowledging the cognitive and operational challenges inherent in mixed-methods research workflows, \citet{zhu2024facilitating} proposed design concepts for notebook environments that more effectively integrate qualitative and quantitative analysis tasks, thereby streamlining hybrid analytical processes.

\subsubsection{Managing Library Dependencies}
Managing dependencies in Jupyter notebooks is crucial not only to ensure their reproducibility, but also to ensure their functionality. To address this challenge, researchers have developed various tools. For example, \textit{SnifferDog} is a dependency management tool that restores notebook execution environments by automatically identifying the required libraries and their compatible versions~\citep{wang2021restoring}. This tool leverages a comprehensive API bank to analyze Python code and make notebooks executable and reproducible. Similarly, \textit{Davos} simplifies dependency management by dynamically installing and updating the correct library versions directly within notebooks, resolving common issues such as version mismatches~\citep{fitzpatrick2022davos}. Another tool, \textit{RELANCER}, employs an automatic technique that restores the executability of broken Jupyter notebooks by upgrading deprecated API calls to non-deprecated ones~\citep{zhu2021restoring}. 

Additionally, a framework that monitors Linux kernel system calls captures specific versions of dependencies to ensure that the computational environment can be precisely recreated across systems, addressing failures caused by missing or incompatible libraries~\citep{wannipurage2022framework}. These tools collectively aim to reduce the manual effort involved in dependency management, improving the reliability of notebooks in diverse computational settings.

\subsubsection{Performance Analysis}
Performance analysis in Jupyter notebooks addresses the efficient execution of notebook code. \citet{werner2021bridging} developed a tool for Jupyter notebooks that measures the execution time and memory usage of individual cells. This tool can identify inefficiencies in runtime or the resource consumption of a notebook to detect performance issues and provide detailed feedback, enabling users to pinpoint resource-intensive areas for refining their code. Extending this idea further, \citet{werner2024jumper} introduced \textit{JUmPER}, which combines coarse-grained performance monitoring with fine-grained instrumentation for interactive HPC workflows. In addition, \textit{JUmPER} leverages parallel marshalling and in-memory communication methods to minimize instrumentation profiling overhead even in distributed computing scenarios. Beyond instrumentation and profiling, other research has explored predictive and benchmarking approaches. For example, \citet{prathanrat2018performance} utilized machine learning models to predict the performance of notebooks running on multi-user JupyterHub servers, helping administrators anticipate potential resource bottlenecks. \citet{faenza2024containerized} examined the impact of containerizing Jupyter notebooks on execution performance, evaluating how the isolation and flexibility provided by containers trade off against any added overhead. For educational use cases, \citet{oden2024integrating} developed an approach to integrate interactive performance analysis into notebooks used in parallel programming courses, allowing students to visualize how code changes affect execution time and resource usage in real time. Additionally, \citet{grotov2025themisto} proposed \textit{Themisto}, a Jupyter-based benchmarking framework to systematically measure and compare the runtime performance of notebook code under different conditions. These contributions complement cell-level profilers by offering higher-level insights into notebook efficiency, thereby guiding users in optimizing their notebooks for better performance.

\begin{tcolorbox}[colback=gray!5, colframe=black, title=Summary of managing computational environment and workflow]
Managing the computational environment and workflow in Jupyter notebooks is critical to ensuring smooth execution. Empirical studies show that notebook workflows typically involve iterative processes, from importing data and editing code to generating output and sharing results~\citep{lau2020design, raghunandan2023measuring}. Tools, such as directed acyclic graphs and unique cell identifiers, have been introduced to help map workflows and manage dependencies between cells~\citep{rehman2019towards}. Researchers developed an efficient computational notebook engine for enabling cell-wise checkpointing~\citep{sato2024multiverse}. Researchers have also explored solutions for adapting notebooks to different computational environments, including tools for migrating cells to distributed platforms such as \textit{Kubernetes}~\citep{duan2023jup2kub} and systems such as \textit{Elastic-Notebook}~\citep{li2023elasticnotebook} for checkpointing and restoration. In addition, managing library dependencies remains a key focus, with tools such as \textit{RELANCER}~\citep{zhu2021restoring}, \textit{SnifferDog}~\citep{wang2021restoring}, and \textit{Davos}~\citep{fitzpatrick2022davos} automating the resolution of API issues and ensuring consistency across environments. These advancements address the unique challenges of Jupyter notebooks, making them more reliable and scalable.
\label{summaryCompEnv}
\end{tcolorbox}

\subsection{Readability of Notebooks}\label{sec:rq2_readability}
The readability of notebooks has two central dimensions: (1)~the readability of the code in the notebook, which focuses on the clarity, organization and quality of the written code~\citep{liu2023refactoring}, and (2)~the readability of the notebook narrative, which emphasizes how effectively the text and the output explain the workflow and analysis~\citep{rule2018ten}. Unlike traditional source code files, notebooks integrate iterative code execution, visual output, and narrative explanations, contributing to a distinctively different in nature narrative-driven coding environment~\citep{rule2018exploration}. Researchers have addressed notebook readability challenges through various practices and tools, with the aim of improving both code and narrative readability.

\subsubsection{Refactoring}\label{sec:rq2_refactoring}
Refactoring helps enhance readability by focusing primarily on improving the clarity and organization of the notebook's code itself. Due to Jupyter Notebook's distinctive characteristics, such as the cell-based structure and the iterative nature of cell-wise execution~\citep{liu2023refactoring}, traditional software refactoring methods may not always be directly applied. Often, notebooks are developed with a focus on completing the final data analysis, and little attention is paid to code refactoring, as users prioritize drawing conclusions and sharing results~\citep{rule2018exploration}. This approach leads to ``messy'' notebooks, which are disorganized and poorly structured, and characterized by scattered cell arrangements, excessive inline outputs, and a lack of modularity~\citep{head2019managing}. Messy code reduces readability, makes notebooks harder to maintain, and complicates tracing the steps that produced the results, which can ultimately discourage sharing and collaboration~\citep{rule2018ten, head2019managing}.

To address these issues, refactoring practices are employed to improve notebook readability and maintainability. Researchers found that the common refactoring practices in notebooks are extracting functions, reordering cells, renaming notebooks, splitting cells, and merging cells~\citep{liu2023refactoring}.  
In another study, \citet{dong2021splitting} expanded this list of refactorings by identifying additional activities such as removing commented blocks, adding comments, and renaming variables. These refactoring practices are not universally applied but vary depending on the authors’ backgrounds (e.g., data scientists versus computer scientists)~\citep{liu2023refactoring}. These also vary from the intended use of the notebook~\citep{dong2021qualitative}. \citet{dong2021qualitative} noticed that notebooks for sharing results with others saw mostly refactorings related to Markdown, whereas, for production notebooks, the most refactorings occurred in the logical code in the transition between notebook and Python file.

In addition, tools such as \textit{nbslicer}, a hybrid static-dynamic slicing tool, were developed to help cleaning and re-executing notebooks more effectively~\citep{shankar2022bolt}. Using backward and forward program slicing, \textit{nbslicer} helps reduce messy code and improves the efficiency of refactoring while addressing challenges such as achieving accurate program slices without excessive overhead or performance issues. These tools and practices collectively aim to improve the readability and maintainability of Jupyter notebooks. Another tool named \textit{ReSplit} helps to split the notebook code cell so that coherent code lines are placed together~\citep{titov2022resplit}. In the first step, the algorithm of that tool suggests merging some of the cells, and in the second step, it suggests extracting specific code fragments from the original cells into new ones~\citep{titov2022resplit}.

Refactoring should not affect the functionality of the code. However, verifying the correctness of a notebook after refactoring can be challenging. For example, when code clones are refactored into functions, users must ensure that the output remains consistent. To address this issue, \citep{sato2022comparing} proposed a verification method by comparing API calls and textual output to ensure that the same API calls with identical parameters produce consistent results.

\subsubsection{Nonlinear Visualization of Notebooks}

\begin{figure}[t]
\centering
\includegraphics[width=0.9\textwidth]{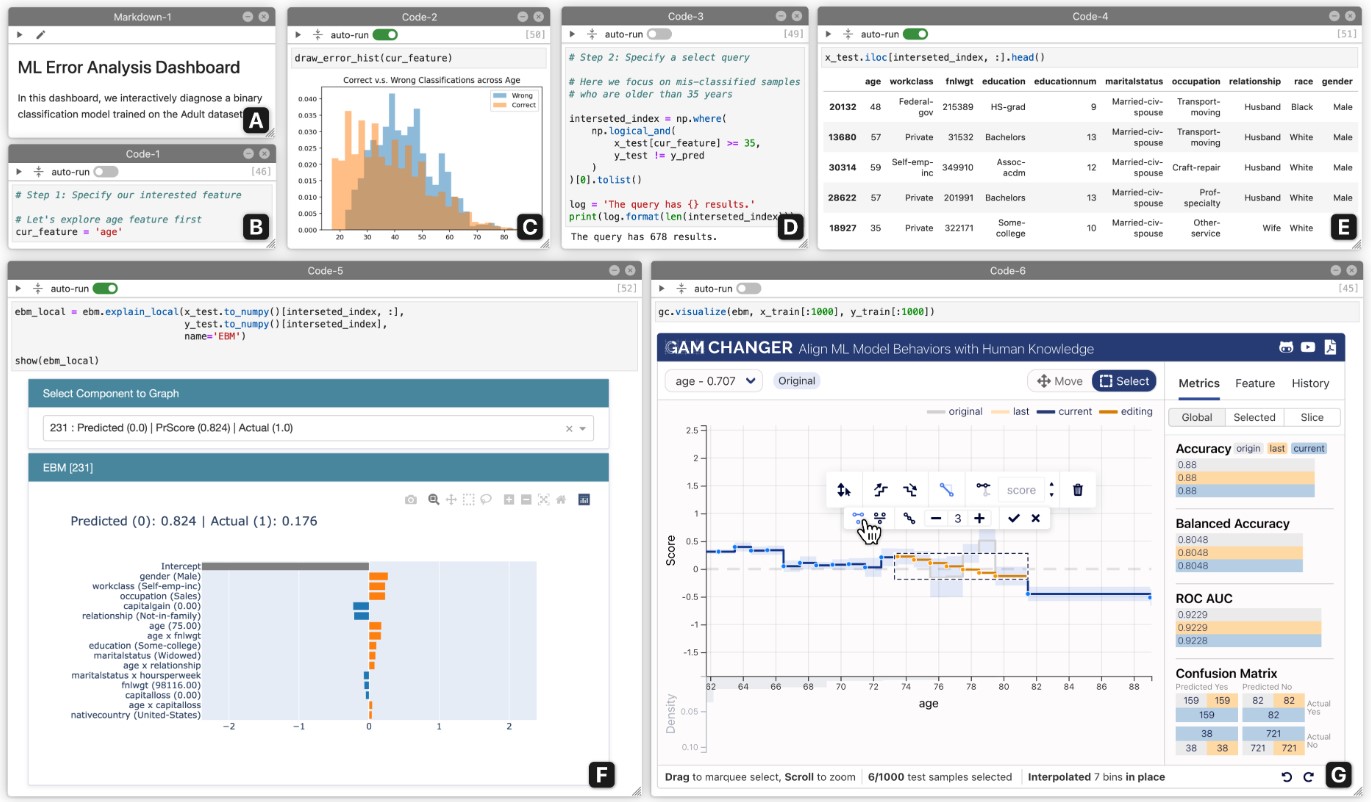}
\caption{A screenshot of the \textit{StickyLand} tool, which helps to place notebook cells in any orientation (not in a fixed top-to-bottom position) (taken from~\url{https://github.com/xiaohk/stickyland})}
\label{fig:stckyland}
\end{figure}

While refactoring primarily focuses on code readability, nonlinear visualization focuses on improving readability through narrative clarity and interactive structuring. Due to the inherent cell-based organization of Jupyter notebooks which follow a linear, top-to-bottom format, the narrative of notebooks can sometimes limit readability. 
Researchers addressed narrative readability by proposing nonlinear organization tools that enable visualization and organization beyond the linear convention. For example, tools such as \textit{StickyLand}~\citep{wang2022stickyland} and \textit{ToonNote}~\citep{kang2021toonnote} allow users to rearrange notebook cells based on their priorities, such as by grouping related sections or highlighting key parts of the workflow. These tools offer new options for organizing, navigating, and displaying notebooks. \textit{StickyLand}~\citep{wang2022stickyland} provides a visual interface where users can freely drag and drop cells to create a customized layout, breaking away from the conventional top-to-bottom order. This allows users to better align the notebook's structure with their thought process or the natural flow of their analysis.
Figure~\ref{fig:stckyland} demonstrates how \textit{StickyLand} transforms the notebook interface into a more intuitive and visually organized workspace.

Building on the idea of enhancing notebook interactivity and structure, \citet{grotov2025evolving} presented the notebook interface as a two-dimensional space, allowing users to spatially organize code cells, detach outputs, and create separate environments for experimentation. Other researchers identified three main 2D cell arrangement patterns in computational notebooks, namely Linear, Multi-Column, and Workboard, highlighting their value for branching and comparative analyses~\citep{harden2022exploring, harden2023there}. Additionally, \citet{rawn2025pagebreaks} proposed \textit{Pagebreaks}, a language construct that introduces scoped regions across multiple notebook cells. Unlike traditional functions, \textit{Pagebreaks} preserve exploratory behaviors such as interleaving code and output or composing programs incrementally, while mitigating confusion caused by global variables. \citet{husak2023slicito} developed a tool to help developers by interactively explore project hierarchies and slice relevant code segments, offering a moldable and task-specific view of complex systems.

Furthermore, \citet{chattopadhyay2023make} explored the cognitive processes involved in understanding computational notebooks, identifying key tasks such as comprehension, mental modeling, and contextual inference. They mapped these tasks to practical design elements, including navigation panels, annotations, and structured titles, and incorporated these elements into \textit{Porpoise}, an interactive overlay tool specifically designed to enhance notebook sense making and navigation. Additionally, \citet{merino2022making} introduced interactive widgets built on an exploring interpreter that visualize execution history and variable states. These widgets allow users to access previous execution across configurations and reduce cognitive load by making program state transitions explicit.

\subsubsection{Accessibility} \label{sec:rq2_accessibility}
Computational notebooks often lack structural features, such as headings, alternative text, and accessible visual outputs, that are essential for blind and visually impaired users to navigate, comprehend, and interact with content via screen readers. To address this issue, \citet{venkatesh2023notably} conducted a large-scale analysis of 100,000 Python-based Jupyter notebooks to identify systemic accessibility barriers in authoring practices, data representation, and publishing workflows. Their findings revealed that most notebooks are structurally inaccessible: 57\% lacked output cells, 43\% contained figures without descriptive alternatives, and semantic elements like headings and tables were underutilized. Additionally, charts and other visual outputs were rarely accompanied by accessible counterparts, such as explanatory text or tabular summaries. To improve accessibility, \citet{nylund2025matplotalt} introduced a Python library called \textit{MatplotAlt}, which enables users to add alternative text (alt text) to Matplotlib figures in computational notebooks. Building on this, we found an urgent need for accessibility aware design in notebook documentation to ensure inclusive practices for notebook developers, including adding alternative text and generating accessible tables alongside visualizations to support screen reader navigation.

\begin{tcolorbox}[colback=gray!5, colframe=black, title=Summary of readability of notebooks]

Notebook readability refers to readability and organization of code and other content within Jupyter notebooks. Researchers have examined refactoring practices to improve code readability, such as reorganizing cells, extracting functions, and splitting or merging cells. Tools such as \textit{nbslicer}~\citep{shankar2022bolt} and \textit{ReSplit}~\citep{titov2022resplit} support these efforts by providing semi-automated solutions to clean up and optimize notebook structures. Moreover, tools such as \textit{StickyLand}~\citep{wang2022stickyland}, \textit{ToonNote}~\citep{kang2021toonnote}, \textit{Pagebreaks}~\citep{rawn2025pagebreaks} facilitate non-linear visualization of notebooks, enabling users to organize cells and better reflect their analytical processes dynamically. Additionally, accessibility remains a critical concern in notebook documentation, with widespread structural barriers limiting screen reader usability and highlighting the need for inclusive authoring practices~\citep{venkatesh2023notably, nylund2025matplotalt}. Collectively, these advancements improve the readability and maintainability of Jupyter notebooks. 
\label{summaryReadability}
\end{tcolorbox}

\subsection{Documentation of Notebooks}\label{sec:rq2_doc}
Research on notebook documentation is crucial because Jupyter Notebook allows combining code with narrative, data visualization, and exploratory analysis in a way that traditional source code files do not~\citep{kluyver2016jupyter}. However, notebook users often pay attention only to the code without creating or updating their documentation~\citep{wang2022documentation}. For our purposes, we define documentation to include any content within Markdown cells and comments in code cells. In this section, we will discuss the key areas that researchers have explored regarding notebook documentation, including document generation, cell header creation, and storytelling. 

\subsubsection{Empirical Studies on Documentation}
Integrating explanatory text in notebook documentation is essential to improve the understandability and shareability of notebooks. These texts make a well-documented notebook that describes workflows, provides context, and explains the purpose of code. Although studies have shown that almost all notebooks (99\%) include at least one Markdown cell~\citep{rule2018exploration}, the quality of documentation inside markdowns is not up to mark~\cite{wang2021makes}. To improve notebook documentation, \citet{wang2021makes} analyzed highly voted notebooks and identified nine guidelines for Markdown content (such as adding sections or subsections) in notebooks. 
Researchers also showed that although using code comments can enhance readability by organizing and annotating the flow of code~\citep{raghunandan2022code}, only a small fraction of notebooks use comments to explain the reasoning (10\%) or expected results (4\%) of the corresponding code cells~\citep{rule2018exploration, rule2018design}. Complementing studies on Markdown and code comments, \citet{wong2025method} found that method names in Jupyter notebooks often lack clarity as embedded documentation, with only 55.57\% starting with a verb and 30.39\% containing abbreviations or acronyms. These naming practices reduce the effectiveness of method names as notebook documentation, underscoring the need for clearer naming conventions in notebook environments.

\subsubsection{Documentation Generation}
Researchers developed several tools to generate notebook documentation. To tackle the challenge of insufficient documentation in data science notebooks, \textit{Themisto}, an AI system to generate code documentation, has been proposed~\citep{wang2022documentation, wang2021graph}. It works in three phases, i.e., retrieving relevant API documentation, generating documentation automatically, and prompting users to add their documentation. Themisto was reported to reduce documentation time, increase user satisfaction, and encourage documentation of previously overlooked code.  
Another tool \textit{HAConvGNN}, uses a hierarchical attention mechanism to focus on relevant code snippets and tokens for accurate documentation generation~\citep{liu2021haconvgnn}. In addition, \citet{schmieg2025enhancing} showed that well-structured notebooks with detailed explanations, dataset descriptions, and visual support enhance knowledge transfer and onboarding in machine learning research. \textit{Cell2Doc} uses ML pipelines to generate documentation for code cells~\citep{mondal2023cell2doc} and \textit{InkSight}~\citep{lin2023inksight} documents insights from charts using sketch-based interactions. 

To enhance the quality and relevance of automatically generated documentation, a CNN-RNN-based approach has been proposed that leverages diverse code metrics such as cell complexity and API popularity~\citep{mostafavi2024beyond}. In addition, incorporating code metrics such as complexity and API usage into deep learning models significantly improves the accuracy of auto-generated documentation, with BLEU and ROUGE scores showing consistent gains~\citep{mostafavi2024can}. Complementing these efforts, a plugin called \textit{InterLink} was introduced to visually link text, code, and outputs in a side-by-side layout, thereby improving the traceability of notebook documentation~\citep{lin2025interlink}.  
Empirical results from \citet{muller2021data} show that in 41\% of the cases, users still need to modify automatically generated notebook documentation.

Storytelling for Jupyter notebooks takes documentation even further by transforming raw analyses into coherent narratives tailored for communication. For instance, \textit{Notable}~\citep{li2023notable} automates the generation of narrative presentation slides directly from the notebooks. Addressing the related challenge of slide creation from disorganized computational notebooks, \citet{wang2024outlinespark} introduced \textit{OutlineSpark} to generate slides through interactive outlining and computational support. Going one step further, \citet{ouyang2024noteplayer} proposed \textit{NotePlayer}, which bridges notebook cells with dynamic video segments by integrating computational engines and LLM-generated narrations, addressing the problem of excessive manual effort traditionally required for detailed tutorial video creation.

\subsubsection{Cell Header Generation}
In notebooks, a cell header is typically a Markdown cell that defines section titles or headings, which helps organize and structure the notebook more effectively. Cell headers often utilize Markdown syntax (e.g., \# or \#\#) to create hierarchical headers, enhancing the readability and navigability. Unlike standard documentation generation, which focuses on providing detailed explanations, inputs, or outputs, cell headers primarily serve to visually structure the flow of the notebook~\citep{venkatesh2021automated}. These headers act as navigational aids, dividing the notebook into logical sections to improve readability and guide users through the workflow.
\citet{venkatesh2023enhancing, venkatesh2024static} introduced an automatic cell header generation tool \textit{HeaderGen}, which analyzes the functions and call graphs of the notebook code to generate appropriate headers for the notebook cells. ~\citet{venkatesh2021automated} published an early prototype of the \textit{HeaderGen} tool that can automatically create headers in Markdown cells using static analysis of the code cells, even without providing specific details about the Markdown text documentation. Extending beyond static analysis, \citet{perez2024flexible} proposed a hybrid approach called \textit{JUPYLABEL}, which incorporated both rule-based heuristics and decision-tree classifiers to classify and label notebook cells to generate headers.

\begin{tcolorbox}[colback=gray!5, colframe=black, title=Summary of documentation of notebooks]
Research on notebook documentation in Jupyter notebooks addressed integrating explanatory text to enhance understandability and shareability. In documentation generation, ML models help automate the process, with tools such as \textit{Themisto}~\citep{wang2022documentation} and \textit{HAConvGNN}~\citep{liu2021haconvgnn} which use ML to generate comprehensive documentation based on code structure and content. Research has explored human interaction with AI-generated text, highlighting the need for human intervention to improve the quality of generated documentation. Cell header generation focuses on automatically generating headers to improve notebook navigability and usability, exemplified by tools such as \textit{HeaderGen}~\citep{venkatesh2023enhancing} and \textit{JUPYLABEL}~\citep{perez2024flexible}. 

\label{summaryDocumentation}
\end{tcolorbox}
\subsection{Testing and Debugging}
\label{sec:testing}
Jupyter notebooks are widely used for prototyping, offering rapid iteration and immediate feedback~\citep{fangohr2020testing}. However, testing in notebooks differs from traditional software testing, as it may require verifying code correctness at the cell level rather than treating the entire notebook as a single executable unit. Due to the lack of standardized testing practices, data scientists have adopted various approaches. For example, some embed test cases within the notebook, while others use separate test notebooks to verify the correctness~\citep{chattopadhyay2020whats}. This section describes studies related to the testing of notebooks, including empirical studies on bug analysis and bug detection.

\subsubsection{Empirical Studies on Testing Notebooks} 
\citet{de2024bug} empirically analyzed the problems and challenges of identifying, diagnosing and resolving bugs in Jupyter notebooks. They presented the first comprehensive study of bugs in Jupyter Notebook projects, analyzing bugs in GitHub repositories and Stack Overflow posts, and interviews with data scientists. The study identifies eight types of bugs that are found in notebooks along with their studies. \citet{shome2024understanding} investigated feedback mechanisms within ML-focused Jupyter notebooks and presented a taxonomy of explicit~(assertions) and implicit~(print statements and last cell outputs) feedback approaches. Their study also provided practical recommendations to encourage explicit testing for minimizing technical debt and improving reproducibility. \citet{wang2024why} analyzed crashes in ML notebooks and categorized predominant exceptions such as \textit{NameError} and \textit{ValueError}. In addition, they reported that some root causes of these exceptions are related to the features of Jupyter Notebook, such as out-of-order cell executions and errors propagating from previous cells.

\begin{table}[t]
\centering
\scriptsize
\caption{Types of bugs in Jupyter notebooks}
\begin{tabular}{p{2.3cm} p{8cm} p{2cm}}
\toprule
\textbf{Bug Type} & \textbf{Description} & \textbf{Study} \\
\midrule

Processing bugs & Issues in memory or computation, i.e., memory overflow, data loss, or performance degradation. & \citep{subotic2022static, negrini2023static, yang2022data, wang2024using, xin2022provenance} \\
Implementation bugs & General coding errors that can cause incorrect results or runtime errors, i.e., syntax, logic, variables, or algorithms-related bugs. & \citep{patra2022nalin, brachmann2020your, grotov2024untangling, grotov2024debug}  \\
Cell defect & Bugs related to notebook cell renderings, i.e., code cells, Markdown, or output interactions. & \citep{fangohr2020testing} \\
Environments and settings bugs & Problems from misconfigured environments or dependencies that interrupt notebook execution, i.e., missing libraries or lack of dependencies. & \citep{brachmann2020your} \\
Kernel bugs & Issues related to the Jupyter kernel, such as crashing, freezing, or failure to start that can interrupt execution. & - \\
Conversion bugs & Errors during notebook conversion from \texttt{.ipynb} file type to other formats, i.e., \textit{Nbconvert} bugs. & - \\
Portability bugs & Failures when running notebooks in different environments, i.e., rendering problems. & - \\
Connection bugs & Failures connecting to external systems like databases or APIs that can block data access and interrupt workflows. & - \\

\bottomrule


\end{tabular}
\label{tab:jupyter_bugs}
\end{table}

\subsubsection{Detecting Bugs}
Detecting bugs or errors in Jupyter Notebook presents unique challenges due to nonlinear workflows and hidden states, distinguishing them from traditional programming environments~\citep{robinson2022error}. Notebook users rely heavily on manual trial-and-error debugging, rather than structured testing practices, with over 70\% of cell re-executions driven by code fixing and cleaning \citep{titov2025observing}. \citet{xin2022provenance} developed a framework to detect anomalies and identify their root causes by combining provenance data with performance metrics. Since Jupyter notebooks do not offer a built-in debugger, the study also presents strategies for debugging Jupyter notebooks by identifying cell workflows and the root cause of errors.
To make the bug detection easier, \citet{tang2025characterising} investigated the underlying root causes of bugs in computational notebooks and identified three dominant sources: incorrect logical implementation, erroneous variable assignments, and file misconfigurations. Complementing these findings, \citet{jiang2025exploring} further emphasized that configuration issues are among the most frequent root causes of notebook bugs, followed closely by improper API usage. Together, these studies highlight the fragile interplay between code logic, environment setup, and external dependencies in notebook-based workflows.

Another bug-detecting study introduced a tool called \textit{Vizier}, which helps debug notebooks by maintaining a complete version history of notebooks, cells, and datasets, while tracking potential errors through fine-grained provenance~\citep{brachmann2020your}. This tool maintains a complete version history for each notebook and allows reproducible data manipulation through a spreadsheet mode~\citep{brachmann2020your}.

Another method to detect potential bugs in notebooks is to focus on inconsistencies in variable assignments, output visualization, or logical operations within the notebook~\citep{patra2022nalin}. For example, output inconsistency occurs when the output of a notebook cell does not match the expected output. An implementation of such an inconsistency-based approach is the \textit{nbval} notebook validation plugin~\citep{fangohr2020testing}. \textit{nbval} detects bugs in Jupyter notebooks by automatically executing each code cell and comparing its output with previously stored results. If there is any deviation, such as an unexpected change in numerical values, formatting differences, or execution failures, \textit{nbval} reports a test failure. \citet{wang2024using} showed that traditional static analyzers frequently fail to detect runtime tensor shape mismatch bugs in ML codes~(particularly TensorFlow) within notebooks. Their study demonstrates how incorporating runtime information improves static analysis capabilities for detecting such common ML-related notebook bugs.

Another type of inconsistency bug is the name-value inconsistency, which occurs when the name of a variable does not accurately reflect its assigned value. Unlike traditional software development environments, notebooks allow for nonlinear execution and frequent variable reuse, increasing the risk of such name-value inconsistencies~\citep{patra2022nalin}.  
For instance, naming a variable \texttt{log\_file} while storing a list of all files in a directory can cause a name-value inconsistency bug, as a developer might assume that \texttt{log\_file} stores a single file name, but it actually contains a list. To avoid confusion, the variable could be renamed to \texttt{log\_files} or \texttt{log\_file\_list}. 
\citet{patra2022nalin} developed \textit{Nalin}, an automated tool for detecting name-value inconsistencies. This tool uses a neural model to predict whether a variable's name and its assigned value are consistent. \textit{Nalin} employs dynamic program analysis alongside deep learning to monitor variable activities throughout the execution process. 

Recently, advancements in LLMs have created promising opportunities for automated error detection and debugging. \citet{grotov2024untangling} performed a detailed analysis of prevalent notebook error patterns and proposed an iterative, LLM-based method for detecting and dynamically resolving errors. Furthermore, \citet{grotov2024debug} developed an LLM-powered AI agent specifically designed to handle stateful and nonlinear notebook debugging cases, demonstrating that agent-based debugging can effectively resolve complex errors in computational notebooks. Their research also highlights the need to balance automation with human oversight, carefully manage interface complexity, and address security considerations (e.g., sandboxing).

\subsubsection{Detecting Data Leakage}
In data science projects, data leakage occurs when information from outside the training dataset is used to develop a model~\citep{subotic2022static}. This issue can arise in Jupyter Notebook if code cells are executed in the wrong order. Although executing cells in an arbitrary sequence may seem harmless, it can lead to data leakage bugs, such as unintentionally sharing information between the training and test datasets. To provide warnings about potential data leakage, a static analysis framework has been developed that uses abstract interpretation for intracell static analysis, ensuring both efficiency and guaranteed termination~\citep{subotic2022static}. Data leakage in ML can be detected early in Jupyter notebooks by turning dataframe operations across cells into a graph. \citet{negrini2023static} showed that building such a graph-based model can capture dataframe operations across all notebook cells in their actual execution order, allowing them to detect data leakage effectively in Jupyter notebooks.
\citet{yang2022data} also tackled data leakage in data science notebooks by developing a static analysis tool that models how data flows across notebook cells. This tool tracks the relationships between datasets, transformations, and model evaluation steps to identify common leakage patterns specific to notebooks, such as the use of test data during preprocessing, the reuse of test data for model selection (known as multi-test leakage), and overlaps between training and test datasets.

\begin{tcolorbox}[colback=gray!5, colframe=black, title=Summary of testing and debugging]
Researchers have focused on developing automated tools designed explicitly for bug detection in notebooks. Those tools can deal with different types of bugs, for example, name-value inconsistencies~\citep{patra2022nalin}, data leakage issues~\citep{yang2022data, subotic2022static}, and output inconsistency~\citep{fangohr2020testing}. Researchers explored LLM-powered solutions for debugging and resolving notebook errors~\citep{grotov2024debug, grotov2024untangling}. However, more research on notebook testing is needed to deal with other bugs such as performance and kernel-related bugs.  
\label{summaryTesting}
\end{tcolorbox}

\subsection{Visualization in Notebooks}\label{sec:rq2_visual}
As an important data analysis outcome, visualization is vital in Jupyter notebooks~\citep{settewong2022visualize, agrawal2022understanding}. It also serves software engineering purposes, as it can be used to validate the correctness of a notebook by allowing comparisons between expected and actual outcomes. Additionally, visualization is crucial for managing and enhancing communication within notebook projects.

\subsubsection{Empirical Studies on Visualization in Notebooks}
Visualization in notebooks can be divided into two main categories: data visualization and workflow visualization. Data visualization refers to the process of displaying the output of a code cell. One of the key features of Jupyter Notebook is the ability to visualize outputs alongside the code cells. Researchers showed that one out of four issues (1,071 out of 4,210) in the studied Jupyter Notebook projects contain at least one visualization-related issue~\citep{agrawal2022understanding}. The study also noticed that visual content is not limited to communicating user interaction design but contains various types of information, such as command-line content or code snippets. \citet{settewong2022visualize} identified nine main reasons for visualizing data when notebook users write code. They analyzed 68 notebooks containing 821 visualizations and categorized them into nine types, showing that the most frequent use cases are for visualizing data distribution (e.g., histograms, box plots) and data frequency (e.g., bar charts, count plots) making these the core tools for understanding and communicating data insights during coding.

In addition, \citet{wootton2024charting} investigated the exploratory data analysis~(EDA) process within Jupyter notebooks and identified distinct temporal and sequential patterns. They proposed quantitative metrics for measuring visualization usage, including revisit counts, representational diversity, and representation velocity. \citet{wang2023supernova} showed that effective visual analytics in notebooks depends not only on what is visualized, but also on how visualizations are designed to fit the notebook environment. Their study identified four key design dimensions that shape the user experience, focusing on how visualizations connect with notebook code, where their data comes from, when they appear, and how easily they can be reused. 

Another type of visualization in notebooks is workflow visualization, which helps users understand and navigate the structure of notebook code. \citet{ramasamy2023visualising} introduced a tool called \textit{MARG}, which visualizes data science notebooks as a graph of decisions, forks, and dead-ends instead of a linear list of cells. This approach reveals the nonlinear nature of real-world notebooks and allows users to trace the workflow more effectively. Their study showed that such visualizations significantly improve the understanding of the workflow of data science notebooks. Expanding on this, \textit{NoteFlow} by \citet{tian2025noteflow} introduced a chart recommendation system for tracing data flows across notebook cells. \textit{NoteFlow} can parse data flow, recommends context-aware charts, and enables forward/backward tracing of visual states, helping users monitor evolving data tables and identify anomalies during exploratory data analysis. Additionally \citet{wenskovitch2019albireo} offered another graph visualization of notebook structure, encoding markdown and code cells as nodes and highlighting variable dependencies.

\subsubsection{Interactive Visualization}
Interactive visualizations in Jupyter Notebook refer to an interactive interface within notebooks that allows users to analyze data through interactive charts. These features greatly enhance the usability and practicality of Jupyter notebooks by facilitating real-time analysis, monitoring, and interaction with code and output~\citep{wu2020b2, kwon2023weedle}. Interactive visualization also provides valuable insights and support informed decision-making by integrating visual analytics directly into the notebook environment. For instance, \citet{scully2024design} developed a notebook-embedded interactive visualization that traces and synchronizes visual components with code cells. Furthermore, \citet{guo2024explainability} developed \textit{bonXAI}, which integrates interactive explainable AI~(XAI) visualizations into Python-based ML workflows within notebooks. \citet{ono2021interactive} presented three core approaches for embedding interactivity, named as, matplotlib callbacks, visualization toolkits like Plotly and Altair, and custom HTML/JavaScript visualizations, which enable users to tailor visualizations to their analytical needs.

Researchers have proposed interactive dashboards to help users explore and analyze their Jupyter notebooks. \citet{kwon2023weedle} developed a customizable interactive dashboard to support data-centric NLP in Jupyter notebooks. Their system includes built-in text transformation operations and various visual analysis features, allowing users to create interactive dashboards seamlessly. Similarly, \citet{wu2020b2} introduced a flexible visualization dashboard that complements traditional code cells, offering users an intuitive interface for data interaction. By importing the necessary library, users can generate visualizations by clicking on dataset columns or creating custom data queries, such as scatter plots. \citet{cai2025jupyter} developed a toolkit for visualizing notebooks' activity, offering real-time dashboards that show cell-level engagement, execution errors, and time-on-task metrics. These ensure real-time updates to code cells based on interactions with the dashboard.

Recent works have also explored extending notebook interaction beyond traditional desktop interfaces. \citet{in2024evaluating} investigated adapting computational notebook interfaces into virtual reality~(VR) and introduced embodied gestures for efficient non-linear exploration to enhance notebook navigation.

\begin{tcolorbox}[colback=gray!5, colframe=black, title=Summary of visualization in notebooks]
Visualization is essential in Jupyter Notebook for data analysis, result validation, and effective communication. Research shows that common applications include visualizing data distributions, statistical measures, and ML workflows~\citep{settewong2022visualize, agrawal2022understanding, tian2025noteflow}. Researchers demonstrated that interactive visualizations enhance usability by enabling real-time interactions with both data and code~\citep{wu2020b2, kwon2023weedle}. Several approaches have been proposed to trace and synchronize visual components with notebook code cells~\citep{scully2024design, guo2024explainability, ono2021interactive}. Additionally, nonlinear workflow visualizations improve the comprehension of complex notebooks, and while established guidelines encourage integration, modularity to improve visual analytics~\citep{wang2023supernova, ramasamy2023visualising, wootton2024charting}.

\label{summaryVisualization}
\end{tcolorbox}
\subsection{Best Practices in Notebooks}
\label{sec:bestPractice}
It is essential to adhere to best practices in Jupyter notebooks to enhance code quality and collaborative efforts~\cite{siddik2023code, grotov2022large}. Unlike conventional software development, Jupyter notebooks are often utilized for exploratory data analysis, often involving frequent adjustments and iterative code execution, resulting in deviations from traditional software development methodologies, such as waterfall or test-driven approaches~\cite{wang2020better}. This section underscores the importance of adhering to best practices in Jupyter notebooks, with a specific emphasis on ensuring code style consistency and addressing the unique challenges of collaborative usage, especially within the realm of ML projects.

\subsubsection{Following Code Style Standards}
One of the prevalent challenges in Jupyter notebooks is the inconsistency in code style arising from the absence of enforced coding standards. Researchers showed that notebooks have 1.4 times more stylistic issues than traditional Python scripts~\cite{grotov2022large}. Research found that Jupyter notebooks are generally larger, but less complex, and tend to use built-in functions much more than scripts and have a higher coupling rate \citep{adams2023comparison}. Another study experimentally demonstrated that Jupyter notebooks are inundated with poor-quality code, such as violations of recommended coding practices, unused variables, and deprecated functions~\cite{wang2020better}. Considering the knowledge-sharing nature of Jupyter notebooks, these poor coding practices might be propagated to future developers.

Lightweight software engineering practices are adopted in Jupyter notebooks to improve clarity, reproducibility, and long-term reuse of research code~\citep{huang2025scientists}. Such practices help balance the need for exploration with the need for organized and understandable code. In their study, \citet{huang2025scientists} identified eight key quality goals, such as clarity, reproducibility, and reusability, and suggested 18 practical tactics to achieve them. Some common tactics include keeping one main task per cell, using markdown cells to explain code and results, and improving or cleaning up code when necessary. Similarly, \citet{cai2025jupyterNotebook} found that notebooks with well-defined and consistently used variables and functions make it easier for others to understand the workflow. This structured approach helps developers and researchers work together more effectively and maintain a smooth flow of collaboration.

Prior work has examined adherence to coding best practices from multiple perspectives. For example, \citet{siddik2023code} evaluated the adherence to the PEP-8 code style guidelines in ML code in Jupyter notebooks. Their findings indicate that ML notebooks generally exhibit lower code quality than non-ML notebooks, with notable discrepancies in how packages and libraries are managed. To address this, a static analysis tool called \textit{Pynblint}~\citep{quaranta2022pynblint, quaranta2024pynblint} was implemented to identify quality issues in Jupyter notebooks.
In addition, \citet{candela2023approach} studied the quality of Jupyter Notebook projects published by GLAM (Galleries, Libraries, Archives, and Museums) institutions within the cultural heritage sector. Their evaluation criteria focused on documentation, code readability, and metadata usage, identifying that these areas need improvement for better traceability and overall quality.

\subsubsection{Best Practices for Collaborative Use}
Researchers showed that it is necessary to develop and validate best practices for collaborative notebook use and the tools required to enforce these practices~\citep{quaranta2022assessing}. \citet{quaranta2022eliciting} introduced a catalog of collaboration-specific best practices for Jupyter Notebook, aiming to improve teamwork, reproducibility, and code quality in data science workflows. The catalog emphasizes practices like using version control, structuring code with modular functions, documenting with Markdown, cleaning and organizing cells, and separating exploratory from production notebooks. It also encourages open sharing and the use of self-contained environments to ensure reproducibility. To manage editing conflicts during real-time collaboration effectively, \citet{wang2024dont} proposed \textit{PADLOCK}, which provides conflict-resolution mechanisms such as cell-level and variable-level access control and parallel cell groups. Their empirical evaluation demonstrated reduced editing conflicts and improved support for diverse collaboration styles. Together, these guidelines and tools support more maintainable and collaborative notebook development.

\begin{tcolorbox}[colback=gray!5, colframe=black, title=Summary of best practices in notebooks]
Unlike traditional software development environments, Jupyter notebooks often prioritize data processing and exploration, which can lead to deviations from established software engineering practices. No studies have identified or explored the ML-specific best practices in notebooks. Building upon the study by \citet{pimentel2021understanding, grotov2022large}, which identified good and bad coding practices, there is a need for an impact analysis of these coding practices.

\label{summaryBestPractice}
\end{tcolorbox}
\subsection{Cell Execution Order}\label{sec:rq2_order}
\label{sec:executionOrder}
Notebook cells can be executed independently and out of order, making it challenging for developers to manage the global variables that notebooks keep persistent in memory~\cite{macke2021automating}. Developers should strive to maintain a linear order of code execution in Jupyter notebooks, as researchers noticed that the nonlinear execution order is one of the main challenges while reusing notebooks~\cite{singer2020notes, macke2021automating}. These non-linear executions in Jupyter notebooks create difficulties for users in understanding and tracing dependencies between code cells, which can cause confusion, errors, and challenges in reproducing results~\citep{brown2023facilitating}. To address this, \citet{brown2023facilitating} introduced Dataflow Notebooks (\textit{DFNBs}), a system that can locate variable definitions and track variable references as explicit data dependencies to overcome out-of-order cell executions. Building on these challenges, \citet{in2025exploring} explored how immersive computational notebooks can support non-linear execution workflows through spatial organization strategies. Spatial organization in computational notebooks refers to how code cells are visually arranged to reflect their logical relationships and execution order. In a controlled study with 20 participants, \citet{in2025exploring} found that users preferred half-cylindrical and quarter-cylindrical layouts to manage notebook execution order.

Developing reproducible ML pipelines can also help address notebook execution order issues. In this context, the ML pipelines in notebooks consist of interconnected code cells where each step of the ML process (such as preprocessing, model training, and evaluation) is represented as a separate code cell~\cite{jiang2022elevating, michael2022keeping}. \citet{jiang2022elevating} developed an approach to determine the correct execution order by identifying and extracting the underlying structure of a notebook by building a labeled dependency graph. In this graph, each cell is represented as a node labeled with a specific ML stage (e.g., data collection, training, evaluation), and edges represent data dependencies between cells. These labeled cells are then reordered to maintain the ML pipeline execution flow.  
A tool named \textit{Ploomber} has been developed to manage and automate the execution sequence of notebook cells~\citep{michael2022keeping}. \textit{Ploomber} transforms notebooks into a structured format, making a Jupyter notebook function like a reproducible pipeline with a single execution flow. This tool allows users to break down large notebooks into smaller, manageable tasks that are connected through a clear execution order. This structured approach helps prevent common issues associated with out-of-order execution, such as undefined variables or incorrect data states.

\begin{tcolorbox}[colback=gray!5, colframe=black, title=Summary of cell execution order]
Nonlinear cell execution in Jupyter notebooks leads to persistent memory challenges, making it difficult for developers to track variable dependencies and ensure reproducibility~\cite{macke2021automating, singer2020notes}. To address this, researchers have developed methods like \textit{DFNBs} for tracking dependencies~\citep{brown2023facilitating}, and tools like \textit{Ploomber}~\citep{michael2022keeping} to automate and streamline notebook workflows. Researchers also developed solutions for reproducible ML pipelines, helping resolve notebook execution order issues~\citep{jiang2022elevating, michael2022keeping}.

\label{summaryCellExecutionOrder}
\end{tcolorbox}

\subsection{AI-based Coding Assistance for Notebooks} \label{sec:rq2_codegen}
Jupyter notebooks offer a convenient way to write and execute code in a single shareable file for exploratory data analysis and insight finding~\cite{kluyver2016jupyter}. However, users with limited coding experience may struggle to participate in the analysis process quickly. AI-based coding assistance for Jupyter notebooks can be a solution for these users by generating code snippets based on various inputs, such as natural language instructions or notebook contexts. Researchers have explored automatic generation of notebook codes with low-code strategies to support data analysis~\cite{chen2023whatsnext, yin2023natural}. For example, \citet{chen2023whatsnext} developed a low-code interaction panel to recommend follow-up questions to guide the next steps in exploratory data analysis~\cite{chen2023whatsnext}. This approach helps users visualize the structure of their data science workflow through a tree-based representation of the notebook cells. 

Notebook users can generate code cells based on natural language instructions. \citet{yin2023natural} proposed an approach to enable users to describe their desired outcome in English and automatically generate the necessary code. However, this solution is designed specifically for tasks involving the \textit{Pandas} library. \citet{huang2024contextualized} further expanded code generation methods by developing \textit{DataCoder}, a dual-encoder model that generates contextualized data-wrangling code cells by encoding textual, code, and tabular data separately to improve accuracy. Users can also generate notebook code cells by interacting with appropriate UI scaffolds. For instance, \citet{cheng2024biscuit} developed \textit{BISCUIT}, a JupyterLab extension leveraging ephemeral user interfaces generated by LLMs, enabling interactive elements~(e.g., sliders, dropdowns) to facilitate more understandable and exploratory ML coding tasks.

AI-assisted code editing in interactive machine learning notebooks faces unique challenges due to their non-linear structure, frequent incremental changes, and rich contextual dependencies. \citet{jin2025suggesting, jin2025learning} demonstrated that the edits on Jupyter notebooks are highly localized and followed incremental maintenance patterns. AI models are increasingly used to assist with these incremental edits, helping developers maintain and evolve their data science workflows more efficiently \citep{jin2025suggesting, jin2025learning}. Notebook editing has been modeled as a sequence-to-sequence task, where AI systems leverage historical revision data and cell-level context to predict realistic and intent-driven modifications. Despite these advances, state-of-the-art LLMs like DeepSeek-Coder-6.7b-instruct \citep{guo2024deepseek} still struggle to handle the complexity and diversity of real-world ML notebooks \citep{jin2025suggesting}.

Notebook users often lose intermediate results during long editing which makes it difficult to reproduce results due to non-linear execution paths. \citet{fang2025large} demonstrated that AI agents like GPT4o can improve the usability of Jupyter notebooks by automatically checkpointing them after editing during execution. The checkpointing approach helps preserve intermediate states, making it easier to recover from failures and understand the computational history of a notebook. Complementing this, \citet{you2025datawiseagent} proposed \textit{DatawiseAgent}, a notebook-centric LLM agent framework that not only integrates execution context and tool usage but also maintains structured agent memory to enhance reproducibility and automate end-to-end data science workflows.

Beyond code generation and modification, studies have emphasized the importance of mediated interactions and enhanced code-cell execution assistance. For educational environments, \citet{george2024notebookgpt} introduced \textit{NotebookGPT}, embedding GPT interactions within Jupyter notebooks to encourage effective prompt usage and reduce the direct copying behavior among students through programmatic mediation strategies. Although these solutions need to connect proprietary closed-sourced AI models, \citet{wandel2025pyevalai} introduced an AI-assisted evaluation system named \textit{PyEvalAI}, which automatically scores Jupyter notebooks using a combination of unit tests and a locally hosted language model. In parallel, efforts have also been directed toward defining new design approaches for seamlessly integrating LLM support into notebook workflows, potentially improving productivity for diverse developer groups~\cite{weber2024computational}. Similarly, \citet{brault2023taming} demonstrated a solution for generating new notebooks by retrieving relevant past experiments based on user-specified problem configurations. \citet{mcnutt2023design} investigated AI-powered notebook assistants that recommend optimal code cell execution pathways by analyzing data dependencies alongside the code itself, highlighting another crucial dimension of notebook assistance beyond generating code snippets.

\begin{tcolorbox}[colback=gray!5, colframe=black, title=Summary of AI-based coding assistance for notebook]
AI-based coding assistance in Jupyter notebooks helps users generate code snippets using natural language instructions or notebook contexts. Researchers have explored a range of techniques, including low-code interaction panels~\citep{chen2023whatsnext, yin2023natural}, context-aware AI models~\citep{huang2024contextualized}, and interactive UI scaffolds~\citep{cheng2024biscuit} to streamline notebook code. Additionally, AI-powered assistants integrated into notebooks facilitate code-cell execution~\cite{mcnutt2023design}, workflow guidance~\citep{brault2023taming}, suggesting edits~\citep{jin2025suggesting, jin2025learning} and enhanced user interactions~\citep{george2024notebookgpt, wandel2025pyevalai} to make notebooks more intuitive for broader audiences.

\label{summaryAIBasedAssistance}
\end{tcolorbox}

\subsection{Supporting other Programming Paradigms}\label{sec:rq2_support}

To support the use of multiple programming languages within a notebook, researchers have proposed polyglot execution environments~\citep{niephaus2019polyjus}. A polyglot notebook system allows code cells to interact directly with data structures and invoke functions or methods from different programming languages~\cite{niephaus2019polyjus, petricek2018wrattler}. This system facilitates direct object sharing and function calling across different programming languages, eliminating the need for data serialization between languages. Hence, a polyglot notebook system can improve the interoperability of various programming languages within a single notebook environment. Extending this line of work, \citet{li2025kernel} introduced \textit{Kernel-FFI}, a system that enables transparent foreign function interfaces in notebooks, allowing seamless invocation of functions across language boundaries without requiring manual data marshaling or kernel switching. \textit{Kernel-FFI} specifically supports object-oriented programming by enabling foreign object referencing and automatic resource management within notebooks.

Furthermore, block-based visual programming within notebooks provides a graphical interface for programming, making the structure and logic of the code visually clear~\cite{verano2022suppose}. Integrating block-based programming into Jupyter notebooks enhances coding practices, simplifies syntax, reduces errors, and lowers the learning curve for non-expert developers. \citet{verano2022suppose} proposed this integration to provide a user-friendly interface for programming and data science tasks, thus facilitating the adoption of computational notebooks across different domains. Similarly, \citet{weber2024extending} developed a multi-paradigm editor within the Jupyter ecosystem that integrates graphical and textual views with automated synchronization, demonstrating enhanced usability, lower workload, and prevention of execution order errors.

\begin{tcolorbox}[colback=gray!5, colframe=black, title=Summary of supporting other programming paradigms]
Supporting multiple programming paradigms in notebooks enhances flexibility by enabling cross-language execution and intuitive visual programming interfaces. For example, polyglot execution environments enable interaction between multiple programming languages within a notebook~\cite{niephaus2019polyjus, petricek2018wrattler}. Additionally, block-based visual programming enhances accessibility by providing a graphical interface, simplifying syntax, and supporting non-expert developers in computational notebooks~\cite{verano2022suppose, weber2024extending}.

\label{summarySupportingOtherProgramming}
\end{tcolorbox}

\subsection{Datasets of Notebooks}\label{sec:rq2_datasets}

\begin{table}[!tbh]
\centering
\caption{Datasets of Jupyter notebooks used in literature}
\label{tab:nbdataset}
\scriptsize
\begin{tabular}{lp{4.5cm}p{4.8cm}p{1.8cm}}
\toprule  
\textbf{Year}        & 
\textbf{Public URL}                                & \textbf{Description}   &
\textbf{Used by}\\ \midrule

2025 & \url{https://doi.org/10.5281/zenodo.14281690} & 48,398 notebook edits derived from 792 Github ML repositories & \citep{jin2025suggesting}    \\
2025 & \url{https://zenodo.org/records/13357570} & 21 notebooks with a log of programmers’
activities in Jupyter Notebook & \citep{nakamaru2025jupyter}    \\
2024 & \url{https://github.com/ISE-Research/DistilKaggle} & 542,051 notebooks from Kaggle with 34 code quality metrics & \citep{mostafavi2024distilkaggle, ghahfarokhi2024predicting} \\
2024 & \url{https://doi.org/10.6084/m9.figshare.26372140} & 297,800 ML notebooks from GitHub and Kaggle to study feedback mechanisms (e.g., assertions and print) & \citep{shome2024understanding} \\
2024 & \url{https://zenodo.org/records/11396773} & 138,376 notebooks from Kaggle used to study data preprocessing practices in ML development & \citep{golendukhina2024unveiling} \\
2024 & \url{https://github.com/Jun-jie-Huang/CoCoNote} & 58,221 code generation examples from GitHub Jupyter notebooks to study data-wrangling & \citep{huang2024contextualized} \\
2024 & \url{https://huggingface.co/datasets/JetBrains-Research/jupyter-errors-dataset} & 10,000 GitHub notebooks containing at least one thrown exception & \citep{grotov2024untangling} \\
2024 & \url{https://zenodo.org/records/15114367} & 64,031 ML notebooks (61k GitHub, 2.7k Kaggle) with 92,542 crashes & \citep{wang2024why} \\
2024 & \url{https://zenodo.org/records/13836922} & 113 notebooks containing 342 recommended cells   &   \citep{aydin2024cellrecommend}  \\
2024  &   \url{https://github.com/bugs-jupyter/empirical-study}   &   105 notebooks from GitHub with 14,740 bug-related commits  &\citep{de2024bug}  \\
2022  &   \url{https://zenodo.org/records/6383115}   &   847,881 notebooks from GitHub to study code structure and style  &\citep{grotov2022large}    \\
2021  &   \url{https://zenodo.org/records/4468523} &  248,761 notebooks from Kaggle (the \textit{KGTorrent} dataset)
   &\citep{quaranta2021kgtorrent, choetkiertikul2023mining, mostafavi2024beyond, ragkhitwetsagul2024typhon, mostafavi2024distilkaggle, golendukhina2024unveiling}    \\
2021  &   \url{https://zenodo.org/records/5109482}   &  267,602 notebooks from Kaggle   &\citep{quaranta2022eliciting}  \\
2021  &   \url{https://ibm.biz/Bdfpk6}    &  3,944 notebooks from Kaggle &\citep{liu2021haconvgnn}   \\
2021  &   \url{https://zenodo.org/records/7109939}    &   470 notebooks from GitHub for analyzing the workflow of data science code    &\citep{ramasamy2023visualising}   \\
2020  &   \url{https://github.com/SMAT-Lab/SnifferDog/blob/master/dataset/all.github.urls}    &   100,000 notebook projects from GitHub, with 507 notebook projects which are executed for reproducibility   &\citep{wang2021restoring}  \\
2020  &   \url{https://zenodo.org/records/3836691}    &   6,000 notebooks from GitHub with code clones  &\citep{koenzen2020code}    \\
2020  &   \url{https://osf.io/9q4wp}  &   2,574 notebooks from GitHub    &\citep{raghunandan2022code}    \\
2020  &   \url{https://figshare.com/s/4c5f96bc7d8a8116c271}   &   200 notebooks from GitHub with meaningful commit histories    &\citep{liu2023refactoring}     \\
2019  &   \url{https://zenodo.org/records/2592524}    &   1.4 million notebooks from GitHub  &\citep{pimentel2019large, oli2021automated, pimentel2021understanding, wang2020assessing, titov2022resplit}   \\
2018  &   \url{https://library.ucsd.edu/dc/object/bb2733859v} &   1.25 million notebooks from GitHub &\citep{rule2018exploration, patra2022nalin, li2021nbsearch}    \\

\bottomrule

\end{tabular}
\end{table}

Publicly available datasets of Jupyter notebooks have been instrumental in advancing research on computational notebooks, enabling large-scale empirical studies on their usage, quality, and evolution. These datasets typically include collections of notebooks from platforms such as GitHub and Kaggle, often annotated with metadata such as commit histories, bug-related information, or workflow analysis. In our systematic literature review, we discovered 21 distinct datasets that are publicly available. Table~\ref{tab:nbdataset} lists all the public datasets of Jupyter notebooks used in the literature, highlighting their publication years, public URLs, the description of sources, and research articles that used them. Among these, 12 were gathered from GitHub, 5 from Kaggle, 2 from both GitHub and Kaggle, and the rest 2 did not mention their data source.

These datasets vary significantly in size and purpose. Some datasets are intended for general research on Jupyter Notebook. For instance, the dataset from \citet{grotov2022large} contains 847,881 notebooks from GitHub, supporting analyses of notebook usage trends. Similarly, the \textit{KGTorrent} dataset includes 248,761 notebooks from Kaggle, offering insights into the structured use of notebooks in data science competitions~\cite{quaranta2021kgtorrent}. Additionally, large-scale collections such as the 1.4 million GitHub notebooks compiled by \citet{pimentel2019large} and the 1.25 million GitHub notebooks by \citet{rule2018exploration} provide a comprehensive view of the Jupyter Notebook platform. \textit{DistilKaggle}~\citep{ghahfarokhi2024predicting, mostafavi2024distilkaggle} extends KGTorrent, containing 542,051 Kaggle notebooks annotated with 34 code quality metrics.

Specialized datasets, such as the one by \citet{de2024bug}, focus on bug-related commits, including 105 Jupyter notebooks from GitHub with 14,740 associated commits, making it highly relevant for debugging and maintenance studies. Similarly, the dataset from \citet{wang2021restoring} emphasizes the reproducibility of notebooks, while the dataset by \citet{koenzen2020code} specifically supports research on code clone detection. Another focused dataset from \citet{ramasamy2023visualising} provides data science code for analyzing the workflow. Furthermore, \citet{grotov2024untangling} built a dataset of 10,000 GitHub notebooks explicitly containing exceptions and error outputs, and \citet{huang2024contextualized} focused on the new task of contextualized data-wrangling code generation leveraging code, data, and text contexts. \citet{aydin2024cellrecommend} introduced \textit{CelRec-DB} dataset, which consists of structured representations of pre-processed Python code cells encoded as semantic vector embeddings (generated using a pre-trained CodeBERT model). Each cell is accompanied by an automatically assigned task label indicating the type of machine learning activity, such as data preprocessing or model training.
To understand data science workflows in Jupyter Notebook environments, another dataset has been created with a detailed log of data science programmers' activities \citep{nakamaru2025jupyter}.
Unfortunately, one of the used datasets~\cite{kallen2021jupyter} was not accessible due to a permission issue and therefore excluded from the table.


Several recent datasets specifically target ML notebooks to address emerging research challenges and trends. For instance, \citet{golendukhina2024unveiling} curated a dataset of 138,376 Kaggle notebooks that contain ML-specific data preprocessing API usage based on the \textit{KGTorrent} dataset~\citep{quaranta2021kgtorrent}. Similarly, \citet{wang2024why} collected 64,031 ML notebooks that contain execution errors (notebook crashes) to support research on debugging practices and error mitigation. \citet{shome2024understanding} constructed a dataset containing 297,800 ML notebooks from both GitHub and Kaggle, including 89.6 thousand assertions, 1.4 million print statements, and 1 million last cell statements. Lastly, \citet{jin2025suggesting} presented 48,398 notebook edits derived from 792 GitHub ML repositories to understand the modification of ML code in notebooks.

\begin{tcolorbox}[colback=gray!5, colframe=black, title=Summary of datasets of notebooks]
Publicly available Jupyter Notebook datasets from GitHub and Kaggle serve various purposes, from analyzing overall notebook usage to studying reproducibility and reusability~\citep{wang2021restoring, koenzen2020code}. Some collections, such as \textit{KGTorrent}~\citep{quaranta2021kgtorrent} and \textit{DistilKaggle}~\citep{ghahfarokhi2024predicting, mostafavi2024distilkaggle}, focus on structured notebook usage, while others, like \textit{CelRec-DB}~\citep{aydin2024cellrecommend}, capture recommended code cells. Additionally, some datasets specifically address machine learning notebooks, supporting research on preprocessing API usage~\citep{golendukhina2024unveiling}, execution errors~\citep{grotov2024untangling}, and debugging practices~\citep{wang2024why, de2024bug}.

\label{summaryDatasetsNotebook}
\end{tcolorbox}

\section{Future Research Directions}
\label{sec:slr:frd}
The future directions of software engineering research on Jupyter Notebook are vast and have great potential to advance the field of data science and computational research. In this section, we outline key future research directions.

\subsection*{Research Direction 1: Improving Jupyter Notebook Code Search through Natural Language Summarization}
As discussed in Section~\ref{sec:rq2_codesearch}, existing code search techniques in Jupyter Notebook primarily use keyword matching or deep learning techniques to search the code directly based on natural language queries~\citep{watson2019pysnippet,li2021nbsearch}. However, these methods can be further improved by incorporating natural language summaries of code functionality into the search process. Future research should explore leveraging LLMs to automate the summarization of individual notebook code blocks, producing concise and descriptive explanations of their purposes. Such summaries not only enrich search accuracy, but also aid users in quickly assessing retrieved results without having to read and interpret the code directly.

\subsection*{Research Direction 2: Building a Broad Set of Notebook-Specific Code Refactoring Tools} 
In Section~\ref{sec:rq2_refactoring}, we discuss different code refactoring tools that offer basic functionalities for notebooks, such as splitting cells~\citep{shankar2022bolt} and reordering notebook cells~\citep{titov2022resplit}. Future research should expand the capabilities of these refactoring tools, addressing common patterns such as the distribution of exploratory workflows across multiple notebooks~\citep{rule2018exploration}. In addition, researchers can target other refactoring features such as detecting and eliminating duplicate code, extracting reusable functions, and restructuring notebooks by merging cells to improve the readability and maintainability of notebooks.

\subsection*{Research Direction 3: Inclusive and Accessible Computational Notebooks}

Despite the widespread adoption of computational notebooks, accessibility remains a largely underexplored research area (discussed in Section~\ref{sec:rq2_accessibility}). Prior studies have shown that accessibility remains a critical challenge in computational notebooks, with large-scale analyses revealing widespread structural barriers that limit usability for blind and visually impaired users~\citep{venkatesh2023notably}. In particular, many notebooks lack essential semantic elements, such as headings and alternative text for visualizations, and rarely provide accessible representations of graphical outputs. While initial tools like \textit{MatplotAlt} enable authors to add alternative text to figures~\citep{nylund2025matplotalt}, future research should move beyond isolated solutions and develop accessibility-aware notebook environments that support inclusive documentation by default. This includes automated detection of accessibility violations, generation of alternative text and accessible tables alongside visual outputs, and integration of accessibility guidance directly into notebook authoring workflows to ensure that accessibility is treated as a first-class quality attribute of computational notebooks.

\subsection*{Research Direction 4: Generating Documentation for Groups of Dependent Code Cells} 
Existing methods for automated documentation generation in notebooks (discussed in Section~\ref{sec:rq2_doc}) focus primarily on the generation of documentation at the individual cell level. For example, \textit{Cell2Doc} generates Markdown descriptions for single cells~\citep{mondal2023cell2doc}, and \textit{HeaderGen} generates structural headings into the notebook~\citep{venkatesh2023enhancing}. However, these single-cell approaches often fail to capture broader contexts or overarching objectives spanning multiple interrelated cells that perform tasks collectively. Since notebook users intentionally cluster related operations into adjacent groups of cells to boost productivity and organization~\citep{wang2022stickyland}, future research should aim to automatically generate context-aware documentation for these clusters. Utilizing analysis techniques such as dependency graph extraction and runtime tracing, researchers can identify groups of cells and automatically produce Markdown summaries.

\subsection*{Research Direction 5: Designing Automated Bug Detection and Remediation Tools Tailored to Notebooks} 
Since bug analysis is crucial to maintaining the quality of Jupyter notebooks (see Section~\ref{sec:testing}), future research could focus on developing automated tools specifically designed for notebook-related bugs. Although some existing approaches address issues such as name-value inconsistencies detection~\citep{patra2022nalin} and data leakage detection~\citep{yang2022data}, further research is needed to tackle more complex bugs, such as cell defects and kernel failure. Future studies should focus on addressing notebook code cell defects, which arise from missing dependencies, incorrect execution order, or undefined variables, as well as kernel crashes, often caused by excessive resource consumption, infinite loops, or faulty package interactions~\citep{de2024bug}. Advanced techniques, including static and dynamic code analysis, execution flow tracking, and visualization-based debugging, can be leveraged to efficiently detect, categorize, and resolve these bugs.

\subsection*{Research Direction 6: Conducting Empirical Studies on the Impact of Best Practices for Coding}
As outlined in Section~\ref{sec:bestPractice}, little is known about the actual influence of following best practices for coding in Jupyter notebooks. Future studies should conduct empirical studies to quantify the effects of following best practices for coding based on categorized effective and ineffective practices~\citep{pimentel2021understanding}. Systematic studies comparing compliant and non-compliant notebooks could provide concrete empirical evidence highlighting the advantages of rigorous adherence to recommended practices. Such evidence would strongly motivate wider adoption of disciplined coding standards among notebook users.

\subsection*{Research Direction 7: Detecting and Resolving Duplicate Execution Numbers in Notebook Cells}
In this SLR, we discuss the studies on execution order in Section~\ref{sec:executionOrder}, but we found no existing studies that explicitly address the issue of duplicate cell execution numbers. This issue occurs when notebook kernels are restarted or when cells are repeatedly executed without proper tracking, resulting in identical cell execution indices. Future studies should develop methods for detecting, tracking and resolving these duplicate execution numbers. Promising solutions include tools for logging, visualizing, and auditing the complete notebook execution history, incorporating timestamps and cell dependency graphs to facilitate clear identification and resolution of duplicate execution numbers.

\subsection*{Research Direction 8: Enhancing AI-Powered Code Generation and Workflow Assistance in Jupyter Notebook}
In Section~\ref{sec:rq2_codegen}, we discuss existing solutions for generating code snippets based on natural language instructions for specific libraries like Pandas~\citep{yin2023natural}. These solutions possess limited scope with a single library and fail to support broader, multi-library, multi-step data science workflow automation. To address these limitations, future research should leverage advanced AI techniques~(e.g., large language models), to improve conversational AI-powered assistants for automated code generation by multi-step workflow automation. For example, interactive low-code panels that recommend follow-up analytical questions and visualize workflow structures~\citep{chen2023whatsnext} could be refined and augmented through continuous conversational AI interactions. By dynamically adapting their recommendations based on real-time user inputs and evolving contexts, these AI-powered assistants could offer more responsive and intuitive exploratory data analysis experiences for notebook users.

\section{Threats to Validity}
\label{sec:slr:threat}
\emph{Internal validity:}
One potential threat to the internal validity of this study is the subjective selection of studies. While we applied clear inclusion and exclusion criteria to select the studies related to software engineering practices in Jupyter Notebook, manual selection may introduce inconsistencies. To mitigate this risk, authors independently labelled the studies and resolved disagreements through discussion. In addition, the data extraction form, which defines the information collected on software engineering practices in Jupyter Notebook, may influence validity. To mitigate this, we refined the form through collaborative discussion between the first and third authors to ensure the form aligned with the research questions and adequately covered the scope. Despite these efforts, some subjectivity is inherent, which we acknowledge as a limitation of the study.

\emph{External validity:}
While most studies analyzed in this review focus on Python notebooks, which may limit generalizability, this focus is justified by the fact that 93\% of notebooks available on GitHub are written in Python~\cite{kallen2021jupyter}. This high percentage highlights the relevance of Python-based research within the broader context of Jupyter Notebook usage. Another concern regarding external validity is that our findings primarily apply to software engineering research on Jupyter notebooks. Future research should expand the scope by investigating other types of research on notebooks to improve the generalizability of the results.

\emph{Construct validity:}
One potential threat to the construct validity of our review is related to the study selection process. We conducted a comprehensive search of all academic articles listed in the IEEE Xplore, ACM, Springer, Elsevier, and DBLP bibliographies, intentionally excluding non-academic sources such as blog posts. Those five digital libraries are widely recognized and frequently used by software engineering researchers to identify relevant articles in systematic literature reviews.

\emph{Conclusion validity:}
Conclusion validity threats can occur due to potential biases in interpreting and synthesizing data, particularly when grouping studies under various software engineering topics. This process required careful interpretation and collaborative discussions between the first and the third authors to finalize the list of software engineering topics. To minimize these threats, we utilized a Trello board\footnote{\url{https://trello.com/}} to categorize the studies according to different software engineering topics and held regular discussions to reach a consensus on the final list.

\section{Conclusion}
\label{sec:slr:conc}

In this systematic literature review, we provide a comprehensive overview of the research landscape of software engineering in Jupyter notebooks. Our review uncovers critical insights and trends that have shaped the field's current state. We utilized a thorough methodology involving identifying, screening, and analyzing 199 relevant research articles (supplementary data are available here\footnote{\url{https://zenodo.org/records/17377280}}). The most important findings of our study are:

\begin{itemize}
    \item Most studies were published in conference venues, indicating the field is rapidly evolving. Interestingly, the most frequent venues were HCI-related instead of core SE venues. The Conference on Human Factors in Computing Systems~(CHI) published 20 studies, whereas the International Conference on Software Engineering~(ICSE) published 7. 
    
    \item Notebook-specific solutions for software engineering issues, such as testing, refactoring, and documentation, are relatively underexplored. For example, resolving duplicate execution numbers, refactoring across notebook clones, and generating group documentation for coherent code cells are future directions derived from our study. 
    
    \item There is a growing need to integrate modern AI-based solutions into Jupyter notebooks to support various software engineering topics, including code search and code generation. By summarizing code functionality and narratives in accessible language, users can more effectively search for the relevance of existing code and facilitate its reuse. Additionally, incorporating advanced AI models can improve code generation and management within Jupyter notebooks.
        
    \item The majority of replication packages are hosted on GitHub, raising concerns about long-term availability and adherence to open science principles, as GitHub repositories can be volatile. In contrast, only 17 studies used permanent repositories such as Zenodo or Figshare, aligning better with open science best practices.

\end{itemize}

This systematic review focuses on recent advances in the area of software engineering applied to Jupyter notebooks. Our analysis and findings can assist researchers and practitioners in addressing challenges such as effectively integrating notebooks into production, ensuring seamless code functionality, and promoting open science principles and reproducibility. By leveraging this knowledge, researchers can work towards optimizing notebooks for greater user-friendliness, reliability, and scalability in real-world applications.


\bibliographystyle{elsarticle-harv}
\bibliography{nbSLR}






\end{document}